\documentclass[journal,compsoc,10pt]{IEEEtran}
\usepackage{graphicx}
\graphicspath{{./_figs/}}
\usepackage{amsmath,amssymb,mathtools}
\usepackage{bm,bbm}
\usepackage{url}
\usepackage{booktabs}
\usepackage{algorithm}
\usepackage{algpseudocode}
\usepackage{multirow}
\usepackage[normalem]{ulem}
\usepackage{amsthm}
\ifCLASSOPTIONcompsoc 
    \usepackage[caption=false,font=footnotesize,labelfont=sf,textfont=sf]{subfig}
\else
    \usepackage[caption=false,font=footnotesize]{subfig}
\fi
\ifCLASSOPTIONcompsoc
  \usepackage[nocompress]{cite}
\else
  \usepackage{cite}
\fi

\theoremstyle{definition}
\theoremstyle{definition}

\algnewcommand\algorithmicinput{\textbf{Input:}}
\algnewcommand\Input{\item[\algorithmicinput]}

\algnewcommand\algorithmicinitialization{0. Initializaton:}
\algnewcommand\Initialization{\item[\algorithmicinitialization]}

\algnewcommand\algorithmicclientupdate{\textbf{1.~Update:}}
\algnewcommand\ClientUpdate{\item[\algorithmicclientupdate]}

\algnewcommand\algorithmicclientprediction{\textbf{2.~Prediction:}}
\algnewcommand\ClientPrediction{\item[\algorithmicclientprediction]}

\algnewcommand\algorithmicclientupload{\textbf{3.~Upload:}}
\algnewcommand\ClientUpload{\item[\algorithmicclientupload]}

\algnewcommand\algorithmicaggregation{\textbf{4.~Aggregation:}}
\algnewcommand\Aggregation{\item[\algorithmicaggregation]}

\algnewcommand\algorithmicbroadcast{\textbf{5.~Broadcast:}}
\algnewcommand\Broadcast{\item[\algorithmicbroadcast]}

\algnewcommand\algorithmicdistillation{\textbf{6.~Distillation:}}
\algnewcommand\Distillation{\item[\algorithmicdistillation]}

\algnewcommand\algorithmiciteration{Steps 1--6 are iterated for multiple rounds}
\algnewcommand\Itaration{\item[\algorithmiciteration]}

\algnewcommand\algorithmicclientselection{2: \texttt{Client Selection:}}
\algnewcommand\ClientSelection{\item[\algorithmicclientselection]}

\algnewcommand\algorithmicdistribution{3: \texttt{Distribution:}}
\algnewcommand\Distribution{\item[\algorithmicdistribution]}

\algnewcommand\algorithmicclientupdatefl{4: \texttt{Client Update and Upload:}}
\algnewcommand\ClientUpdateFL{\item[\algorithmicclientupdatefl]}

\algnewcommand\algorithmicclientuploadfl{5: \texttt{Client Upload:}}
\algnewcommand\ClientUploadFL{\item[\algorithmicclientuploadfl]}

\algnewcommand\algorithmicaggregationfl{5: \texttt{Aggregation:}}
\algnewcommand\AggregationFL{\item[\algorithmicaggregationfl]}

\algnewcommand\algorithmicseven{7:}
\algnewcommand\Seven{\item[\algorithmicseven]}

\algnewcommand\algorithmicsix{6:}
\algnewcommand\Six{\item[\algorithmicsix]}

\title{
    \fontsize{20}{24}\selectfont
    Distillation-Based Semi-Supervised Federated Learning \\
    for Communication-Efficient Collaborative Training\\
    with Non-IID Private Data
}

\begin{document}
\author{
    Sohei~Itahara,~\IEEEmembership{Student~Member,~IEEE,}
    Takayuki~Nishio,~\IEEEmembership{Senior~Member,~IEEE,}
    Yusuke~Koda,~\IEEEmembership{Student~Member,~IEEE,}
    Masahiro~Morikura,~\IEEEmembership{Member,~IEEE,}
    and~Koji~Yamamoto,~\IEEEmembership{Senior~Member,~IEEE,}
    \IEEEcompsocitemizethanks{
        \IEEEcompsocthanksitem The authers are with
        Graduate School of Informatics, Kyoto University, Kyoto 606-8501, Japan \protect\\
    E-mail: nishio@i.kyoto-u.ac.jp
    }
}

\IEEEtitleabstractindextext{
\begin{abstract}
    This study develops a federated learning (FL) framework 
    overcoming largely incremental communication costs due to model sizes in typical frameworks without compromising model performance.
    To this end, based on the idea of leveraging an unlabeled open dataset, 
    we propose a distillation-based semi-supervised FL  (DS-FL) algorithm that exchanges the outputs of local models among mobile devices,
    instead of the model parameter exchange employed by typical frameworks.
    In the proposed DS-FL, the communication cost relies only on the output dimensions of the models and does not scale up according to the model size.
    The exchanged model outputs are used to label each sample of the open dataset, creating an additionally labeled dataset.
    The newly labeled dataset is used for further training the local models,
    and model performance is enhanced owing to the data augmentation effect.
    We further highlight that in the proposed DS-FL,
    the heterogeneity of the devices' dataset leads to the ambiguity of each data sample, lowering the training convergence.
    To prevent this, we propose entropy reduction averaging, where the aggregated model outputs are intentionally sharpened.
    Moreover, the extensive experiments conducted show that DS-FL reduces the communication costs up to 99\% relative to those of the FL benchmark while achieving similar or higher classification accuracy.
\end{abstract}
\IEEEpeerreviewmaketitle
}

\maketitle
\IEEEdisplaynontitleabstractindextext

\IEEEraisesectionheading{\section{Introduction}}
\IEEEPARstart{F}{ederated} Learning (FL)~\cite{FL_survey, park2019wireless, lim2020federated, FedAve} is an emerging machine learning (ML) framework to perform data-driven analysis or decision making, leveraging privacy-sensitive data from mobile devices.
Typically, in FL, mobile devices collaboratively train their local ML model through the periodical exchange and aggregation of ML model parameters or gradients at central servers rather than exchanging their raw data.
Thus, FL differs from typical ML in which raw data is acquired and stored in central servers where the private data of the mobile users can be exposed.
Owing to the privacy advantage, FL can be applied to model training tasks with privacy-sensitive data.
For example, Google-keyboard query suggestions from the typing history of mobile users, containing privacy-sensitive information such as the credit card information of the user or login credentials~\cite{Google_FL}.

Despite the benefits of FL, relying on distributed mobile devices generally poses new inconveniences related to communication efficiency~\cite{FL_survey}.
Specifically, the periodical model parameter exchange in typical FL entails communication overhead that scales up according to the model size.
This prohibits the use of large-sized models, particularly when the mobile devices are connected to wireless networks while competing for limited radio resources, which can be a crucial bottleneck for building practical ML models.
Hence, an FL framework that can be scalable according to the size of the models in terms of communication efficiency is required.

Motivated by the inconvenience mentioned above, we aim to answer the following question: 
\textbf{
    How should an FL framework be designed scalable according to the model sizes in terms of communication efficiency while achieving model performance comparable to that of the benchmark FL designed in \cite{FedAve}?
}
Concisely, our answer is leveraging an unlabeled open data shared among the clients to enhance the model performance of model output exchange methods.

To achieve the scalability of the communication overhead,
we leverage the principle of FL with \textit{model output exchange} instead of model parameter exchange.
Here, the exchanged model outputs of mobile devices are named ``local logit'' instead of model parameters.
The local logits are ensembled in a central server into a ``global logit'' that is regarded as \textit{teacher} knowledge.
Moreover, this knowledge is transferred into local models as \textit{students}.
In the model output exchange, communication overheads depend only on the model output dimension, which is often substantially smaller than the number of model parameters and cannot scale up regarding model sizes.

Hence, answering the above question boils down to designing an FL framework with model output exchange to achieve similar performance to the benchmark FL designed in~\cite{FedAve}.
Although the FL with model output exchange is available in the literature, this task remains nontrivial because of the several challenges we describe below.

\vspace{3pt}\noindent \textbf{Leveraging unlabeled data towards performance similar to that of benchmark federated learning under non-IID.}\quad
Typical FL with model output exchange termed federated distillation (FD)\cite{FD, oh2020mix2fld, ahn2019wireless, ahn2020cooperative} can achieve scalability of the model size; however, these methods provide poor models in general.
In FD, each mobile device re-trains local model based on both the local labeled data and the global logits.
However, under non-IID data distribution, where the local dataset of the mobile user does not represent the population distribution, the global logit retains similar information to the local labels already attached to each mobile device.
Hence, this re-training can be almost identical to local model training.
(see ``1.~Update'' and ``6.~Distillation'' procedures in Fig.~\ref{fig:DS-FL_process}~(b) in Section~\ref{sec:proposed_method} for detail).
Thus, the model performance trained in FD is worse than that in an FL benchmark exchanging model parameters~\cite{ahn2019wireless}.
Moreover, achieving similar performance to the FL benchmark is challenging.

Our key idea is to share the unlabeled open data among mobile devices and leverage the data for distillation to overcome this challenge.
In this regard, we propose a novel FL framework-exchanging model outputs named distillation-based semi-supervised FL (DS-FL).
Unlike FD, in the proposed DS-FL, teacher knowledge is used to label the unlabeled data instead of local data already labeled.
The procedure creates novel labeled data.
Subsequently, through the re-training of the local models based on the dataset, the model performance is enhanced due to data augmentation effects (see ``6.~Distillation'' procedure in Fig.~\ref{fig:DS-FL_process}~(c) in Section~\ref{sec:proposed_method} for detail).
The ML experiments show that the proposed DS-FL achieves similar or higher performance compared to that of the FL benchmark with model parameter exchange while reducing communication overheads. To the authors’ best knowledge, our approach has not been considered before.

\vspace{3pt}\noindent \textbf{Logit aggregation towards faster convergence.}\quad
In DS-FL, uploaded logits are aggregated into a global logit, providing an inferred probability of each unlabeled data belonging to a particular class.
However, owing to non-IID data distributions, each uploaded logit exhibits heterogeneity.
Moreover, the aggregated global logit may represent \textit{ambiguous knowledge} where the global logit exhibits a high entropy.
Thus, indicating that the global logit provides incorrect knowledge regarding which class each sample of the unlabeled data pertains, resulting in slower convergence of model training.

Motivated by the challenge mentioned above, we propose entropy reduction aggregation (ERA) that intentionally reduces the global logit entropy.
The ML experiments show that DS-FL with ERA leads to faster convergence while achieving higher classification accuracy than an FL benchmark under non-IID distribution.
Another positive consequence of introducing ERA is to enhance the robustness against several attacks of malicious users uploading corrupted local logits, which is also verified in the ML experiments.

\subsection{Our Contributions}
The contributions of this paper are summarized as follows:
\begin{itemize}
    \item 
    Based on the fundamental idea of leveraging unlabeled open data, we propose an FL framework named DS-FL, which is communication-efficient and achieves a high model performance.
    In more detail, DS-FL exhibits a model performance similar to that of an FL benchmark even under non-IID data distributions while achieving scalability according to model size in terms of communication efficiency.
    With the exchanged model outputs, the unlabeled data acquire labels, based on which each local model can be further trained.
    Therefore, model performance is enhanced due to the data augmentation effects.
    Moreover, DS-FL exchanges model outputs instead of model parameters, where the communication overhead cannot scale up according to the model size.
    The ML experiments show that DS-FL reduces the communication overheads by up to 99\% while achieving similar model performance relative to the FL benchmark even under non-IID data distributions.
    \item 
    We develop a novel model-output aggregation method, named ERA, which is robust against the heterogeneity of the uploaded model output due to non-IID data distributions that result in a slow training convergence.
    First, we highlight that the heterogeneity of the uploaded model output leads to higher entropy of the aggregated model outputs, which is the principal cause of the challenging slower convergence.
    Hence, the key idea behind introducing ERA is to reduce the entropy of the global logit intentionally before distributing it into mobile devices.
    The ML experiments verify that ERA achieves faster convergence than a baseline considering simple averaging (SA) aggregation, thus, reducing the cumulative communication cost by up to 75\%.
\end{itemize}

The scope of this paper is to design an FL framework, satisfying two requirements: to acquire scalability to the model size and perform well under non-IID data.
The existing FL framework satisfying the former requirement is FD, 
whereas FD's performance is substantially low under non-IID data.
Thus, designing an FL framework satisfying these two requirements contributes to the body of knowledge.
Additional requirements, such as performing well under unbalanced and/or massively distributed data, are out-of-scope of the study.

In parallel with and independent of this work, a similar concept of sharing an unlabeled dataset among mobile devices has presented in \cite{chang2019cronus}. However, the study relies on a different motivation; it is based on enhancing attack robustness from malicious mobile devices.
Our initial study~\cite{itahara2019astudy} was presented at a domestic conference in parallel to \cite{chang2019cronus}.
Meanwhile, unlike \cite{chang2019cronus},
in this study, we investigate how DS-FL can achieve similar performance to FL benchmark under non-IID distributions in an efficient communicational manner.
More specifically, we provide a novel logit aggregation method, i.e., ERA, enhancing communication efficiency under a non-IID dataset.
Moreover, we provide a comparison between DS-FL, FL benchmark, and FD through ML experiments using non-IID datasets, which were not considered in \cite{chang2019cronus}.

\subsection{Related Work and Paper Organization}
\label{sec:2_related_works}

\vspace{3pt}\noindent\textbf{Federated learning.}\quad 
FL~\cite{FL_survey, park2019wireless, lim2020federated, FedAve,liu2020federated} is a distributed learning framework enabling ML model training using privacy-sensitive datasets of mobile devices while keeping all the datasets local.
In typical FL, mobile devices collaboratively train their local ML model through the periodical exchange and aggregation of ML model parameters or gradients at central servers rather than exchanging their raw data.
Thus, the central server and mobile devices obtain a qualified ML model, trained using the private dataset on mobile devices without exposing the privacy-sensitive data.

\begin{figure*}[t!]
    \centering
    \subfloat[Benchmark~1: Federated Learning with model parameter exchange~\cite{FedAve}.]{\includegraphics[width=0.31\textwidth]{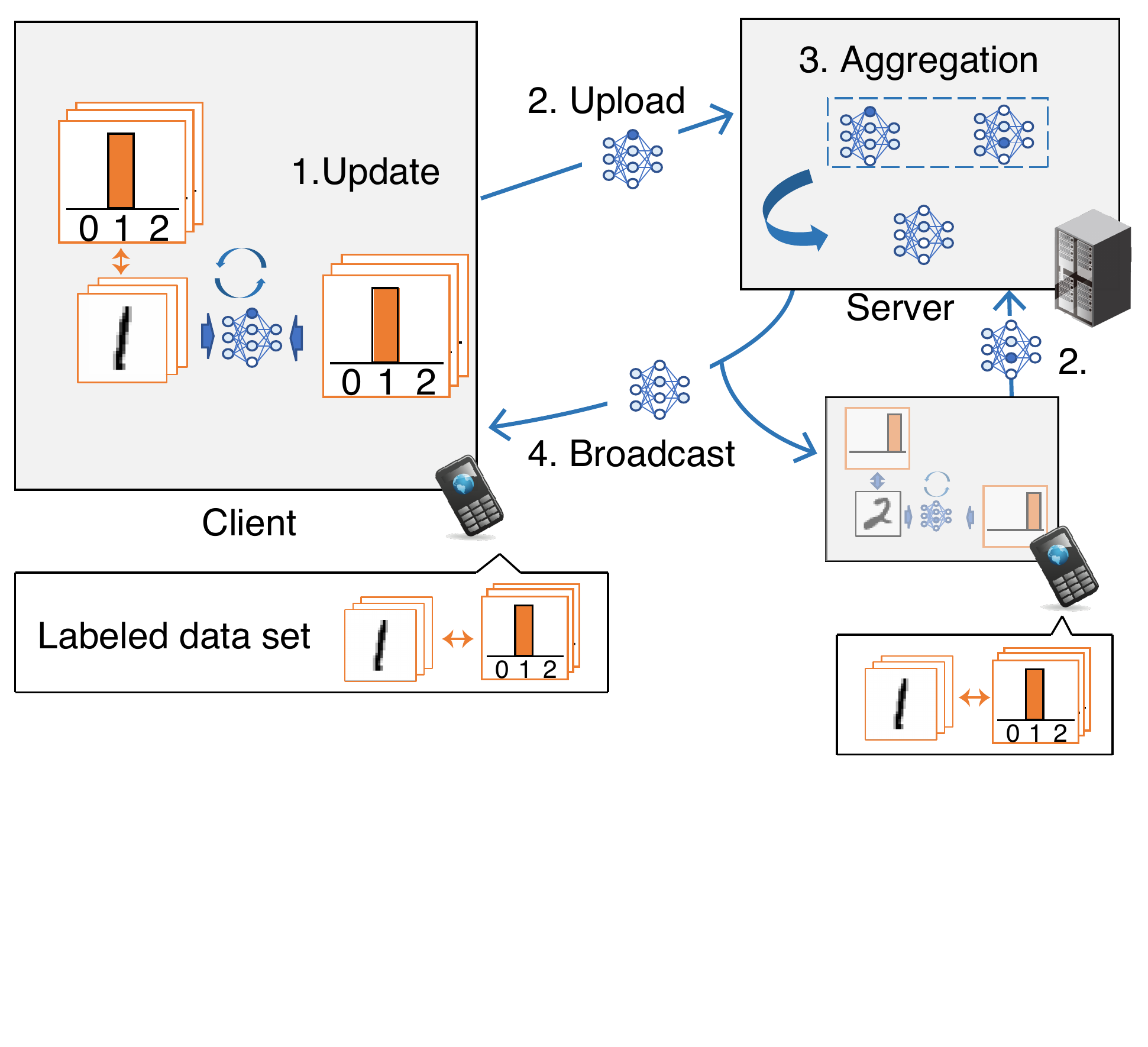}}\hspace{1em}
    \subfloat[Benchmark~2: Federated Distillation~\cite{FD}.]{\includegraphics[width=0.31\textwidth]{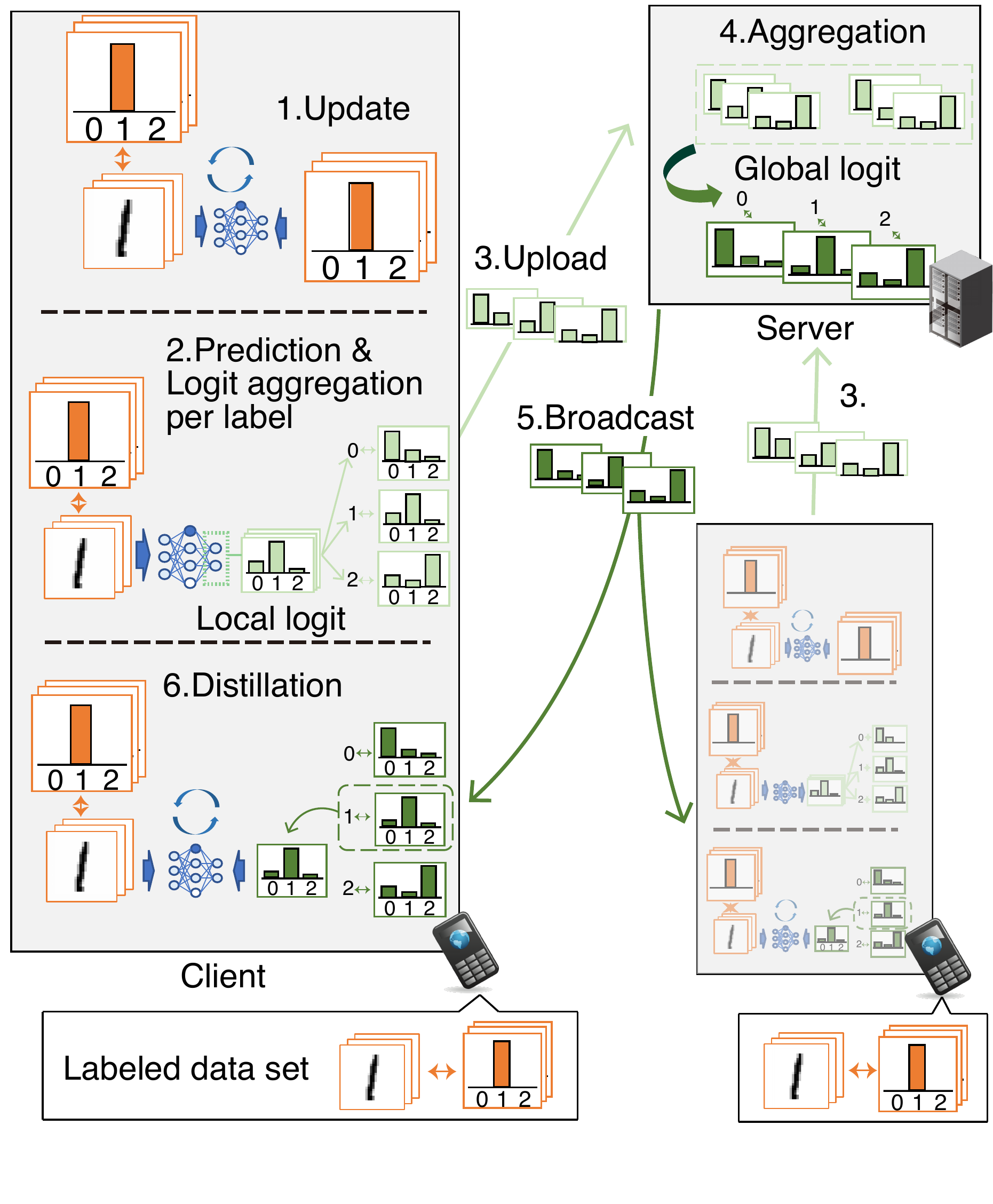}}\hspace{1em}
    \subfloat[Proposed: \textbf{Distillaion-Based Semi-Supervised Fedetrated Learning}.]{\includegraphics[width=0.31\textwidth]{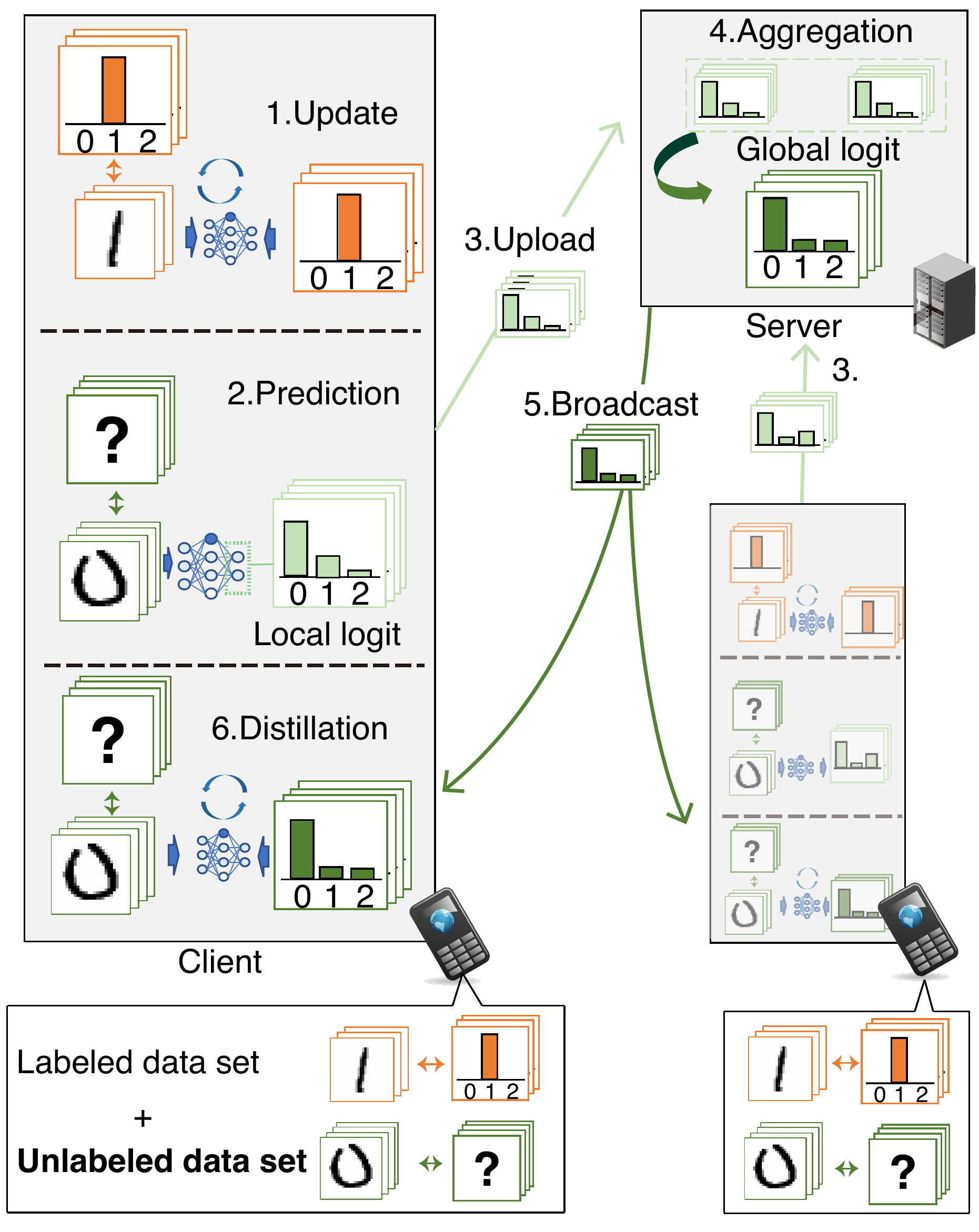}}
    \caption{
        Operational structures for benchmark schemes and proposed DS-FL.}
    \label{fig:DS-FL_process}
\end{figure*}

\vspace{3pt}\noindent\textbf{Communication-efficient federated learning with model parameter exchange.}\quad
Several studies have focused on reducing the communication cost in model parameter exchanges in FL.
For example, the initial study that proposed FL addressed this problem by increasing the number of local model updates and exchanging the model parameters less frequently \cite{FedAve}.
In this approach, network traffic reduces drastically compared to that of algorithms considering iterating local model update and model parameter exchange alternately\cite{chen2016revisiting}.
An alternative strategy is to limit the number of participating mobile users by selecting the users satisfying the stringent requirement for model update time\cite{nishio2018client, dhakal2019coded}. The network traffic can be reduced relative to all users participating in the FL.
Other approaches generating an additional labeled dataset,
distributed and used on the clients, are~\cite{zhao2018federated,FD}.
Indeed, these approaches improve communication efficiency and model performance.
Another stream of research proposed the model compression to reduce the communication cost required for the model parameter exchanges, which can be performed via several strategies, such as low-rank representation~\cite{konevcny2016federated}, model parameter quantization~\cite{TWN, signsgd, konevcny2016federated}, neural network pruning~\cite{itahara2020lottery}, update reuse~\cite{chen2018lag}, parameter sparsification~\cite{STC_FL, konevcny2016federated}, and the Chinese remainder theorem~\cite{xianglong2020privacy}.
However, these studies relied on the model parameter exchange, where the communication-overhead increases proportionally to the model size.
Unlike these studies, we aim to design an FL framework scalable for model size in terms of communication efficiency by designing an FL framework exchanging model outputs instead of the entire model parameters.

\vspace{3pt}\noindent\textbf{Distributed training with model output exchange in data center application over shared dataset.}\quad
Co-distillation (CD)~\cite{co_distillation, zhang2018deep} is a basic distributed learning method with model output exchange.
In CD, distributed ML models are trained over a shared labeled dataset.
Subsequently, each local logit from the trained distributed ML models is exchanged and aggregated into global logit acting as teacher knowledge.
Finally, the teacher knowledge is transferred into each distributed ML model acting as a student by re-training the ML model using the global logit.
Note that this framework is an extension of the knowledge distillation presented in \cite{distillation} for multi-party training.
Another distributed learning method similar to CD, named private aggregation of teacher ensembles (PATE), was proposed in \cite{papernot2016semi, papernot2018scalable}, where the teacher knowledge transfer is performed over the unlabeled dataset instead of the labeled dataset.
While both CD and PATE improve each distributed ML model in a communication-efficient manner, the assumption that each distributed ML model is trained over a shared labeled dataset is suitable for parallel model training in the data center.
However, the assumption is not suitable for model training with data generated on mobile devices.
Unlike these training methods, we design an FL framework with model output exchange, enabling model training with mobile device-generated data subjected to challenging non-IID data distributions, as discussed in the previous section.

\vspace{3pt}\noindent\textbf{Federated learning with model output exchange over mobile device-generated dataset.}\quad
FD is proposed in \cite{FD, oh2020mix2fld, ahn2019wireless, ahn2020cooperative} as an FL framework with model output exchange that trains ML models considering mobile device-generated dataset.
Unlike CD and PATE that trains distributed ML models using a shared dataset, each mobile device trains each ML model using a local dataset, enabling ML model training with mobile device-generated data.
While FD performs well when the mobile device-generated data is identically and independently distributed, FD exhibits lower performance than the FL benchmark with model parameter exchange in non-IID data distributions.
This was experimentally verified in \cite{FD, oh2020mix2fld, ahn2019wireless, ahn2020cooperative} and the experiments presented in Section~\ref{sec:experiment}.
To fill this gap, we design an FL framework with model output exchange achieving similar or higher performance than previously proposed approaches even when subjected to non-IID data distributions.

\vspace{3pt}\noindent\textbf{Semi-supervised federated leanining.}\quad
A few semi-supervised FL frameworks~\cite{kang2020fedmvt,jin2020survey ,zhang2020benchmarking}, using both unlabeled and labeled data, have been proposed.
The work in~\cite{kang2020fedmvt} aimed to improve vertical FL (VFL), which builds a machine learning model based on vertically portioned data (e.g., multiview images).
In this setting, the model parameters or gradients are not generally uploaded, and hence, the communication cost is negligible.
Being different from~\cite{kang2020fedmvt}, as in the benchmark frameworks~\cite{FedAve,FD}, this study considers horizontal partitioned data and addresses the communication costs for uploading model parameters or gradients.
Other works~\cite{jin2020survey,zhang2020benchmarking} considered training the model using labeled data on the server and unlabeled data on the clients. 
While the communication costs in~\cite{jin2020survey} and~\cite{zhang2020benchmarking} increase with the model size, we aim to achieve communication efficiency scalable to the model sizes.

\vspace{3pt}\noindent\textbf{Paper organization.}\quad
The remainder of this paper is organized as follows:
Section~\ref{sec:proposed_method} describes the proposed DS-FL framework.
Section~\ref{sec:make_pseudo_label} presents the proposed logit aggregation method.
Section~\ref{sec:experiment} provides the experimental results where a comparison between DL-FL, FL benchmark, and FD is presented.
Finally, the concluding remarks are presented in Section~\ref{sec:conclusion}.

\section{Distillation-Based Semi-Supervised Federated Learning Method}
\label{sec:proposed_method}
We propose a DS-FL aiming at communication efficiency while achieving similar or higher model performance than several benchmarks.
We summarize the benchmark schemes and the proposed DS-FL in Fig.~\ref{fig:DS-FL_process}, detailed as follows.

\subsection{Benchmark~1. Federated Leaning with Model Parameter Exchange}
\label{ssec:BL_FL}
In the FL with model parameter exchange, mobile users, called \textit{clients} as per terminology, collaboratively train ML models while exchanging the model parameters, as shown in Fig.~\ref{fig:DS-FL_process}~(a).
Specifically, the training procedure in FL with model parameter exchange includes four steps: ``1.~Update,’’ ``2.~Upload,’’ ``3.~Aggregation,’’ and ``4.~Broadcast.’’
These steps follow an iterative process until training converges.
In ``1.~Update'' step, every client trains its local ML model using its own labeled dataset.
The ``1.~Update'' step is common to the DS-FL. 
Subsequently, in ``2.~Upload'' step, the clients share the model parameters with a remote server.
Finally, the server aggregates the uploaded model parameters to build the global model in the ``3.~Aggregation'' step and broadcasts the parameters of the global model to the clients in the ``4.~Broadcast'' step.

The detailed procedure of FL is described below.
In the following, we consider that each client $k = 1, 2, \dots, K$ holds the labeled private dataset $(\bm{d}^{\mathrm{p}}_{i, k}, \bm{t}_{i, k})_{i = 1}^{I_k}$, where $\bm{d}^{\mathrm{p}}_{i, k}$ represents the vectorized input samples.
Moreover, $I_k$ denotes the number of samples in the labeled dataset.
Considering $N_\mathrm{L}$ as the number of objective class,
the term $\bm{t}_{i, k} = [t_{i, k, 1}, \dots, t_{i, k, N_{\mathrm{L}}}]^{\mathrm{T}}$ is the vectorized form of the label attached to the sample $\bm{d}^{\mathrm{p}}_{i, k}$ and is in the one-hot representation, wherein the element $t_{i, k, n}$ equals 1 if the $n$th label is the ground-truth and 0 otherwise.
For shorthand notation, let $N_{\mathrm{S}}\times I_k$  matrix $\bm{D}_k^{\mathrm{p}}$ denote the concatenation of $(\bm{d}^{\mathrm{p}}_{i, k})_{i = 1}^{I_k}$, where $N_{\mathrm{S}}$ represents the dimension of input samples.

\vspace{.3em}\noindent\textbf{1.~Update.}\quad
In this step, each client updates its model with its private dataset based on the stochastic gradient descent algorithm~\cite{Goodfellow2016Deep}.
The initial values of the model $\bm{w}_0$ is distributed from the server before each ``1.~Update'' step.
Specifically, the model parameter is updated as follows:
\begin{align}
	\label{equ:model_update}
	\bm{w}_k \gets \bm{w}_0 - \eta \nabla \phi (\bm{D}_k^{\mathrm{p}}, \bm{T}_k\,|\,\bm{w}_0),  
\end{align}
where $\phi(\cdot, \cdot\,|\,\bm{w}_k)$ denotes the loss function that is minimized in this step.
The loss function is exemplified in classification problems by the cross-entropy. In this case, $\phi (\bm{D}_k^{\mathrm{p}}, \bm{T}_k\,|\,\bm{w}_k) $ is given as follows:
\begin{multline}
    \label{equ:model_update_second}
    \!\!\!\!\!
    \phi (\bm{D}_k^{\mathrm{p}}, \bm{T}_k\,|\,\bm{w}_k)\!\!=\!\! 
    -\!\!\sum_{i\in\mathcal{I}^{\mathrm{rd}}_k}\sum_{n \in \mathcal{N}_{\mathrm{L}}} t_{k, i, n}\log F_{n}(\bm{d}^\mathrm{p}_{i, k}\,|\,\bm{w}_k),\!\!\!
\end{multline}
where $F_{n}(\cdot\,|\,\bm{w}_k)$ denotes the $n$th element of $F(\cdot\,|\,\bm{w}_k)$.
In~\eqref{equ:model_update} and~\eqref{equ:model_update_second}, $\eta$ represents the learning rate, $\mathcal{N}_{\mathrm{L}} \coloneqq \{1, 2, \dots, N_{\mathrm{L}}\}$, and $\mathcal{I}^{\mathrm{rd}}_k\subset \{1, 2, \dots, I_k\}$ is the index set of the minibatch that is randomly sampled from $(\bm{d}^{\mathrm{p}}_{i, k})_{i = 1}^{I_k}$.
The update procedure is an iterative process until a terminating condition, such as convergence or a predefined number of iteration times, is satisfied.

\vspace{.3em}\noindent\textbf{2.~Upload.}\quad
The updated model parameters $\bm{w}_k$ or its gradients $\bm{g}_k = \bm{w}_k - \bm{w}_0$ are uploaded from each client to the server.

\vspace{.3em}\noindent\textbf{3.~Aggregation and 4.~Broadcast.}\quad
The server aggregates the uploaded models from clients to update the global model $\bm{w}_0$, as follows:
\begin{align}
	\bm{w}_0 =  \sum_{k=1}^{k=K} \frac{I_k}{I} \bm{w}_k,
\end{align}
where $I = \sum_{k=1}^{K}I_k$.
Subsequently, the server broadcasts the global model to all the clients via multicast channels.
These procedures are iterated for a finite number of rounds.

\subsection{Benchmark~2. Federated Distillation}
\label{ssec:BL_FD}
Fig.~\ref{fig:DS-FL_process}~(b) shows the process of FD~\cite{FD}, i.e., one of the FL algorithms with model output exchange, where clients share per-class logits instead of model parameters.
In the FD, each client treats itself as a student, while aggregated logits act as teachers, where each local client model is trained using the aggregated logits.
The specific procedure in FD consists of the following six steps: ``1.~Update,’’ ``2.~Prediction \& Logit aggregation per label,’’ ``3.~Upload,’’ ``4.~Aggregation,’’ ``5.~Broadcast,’’ and ``6.~Distillation,’’ as shown in Fig.~\ref{fig:DS-FL_process}~(b).
The differences between FD from the benchmark~1 are in the second and sixth steps.
After the ``1.~Update'' step, using the trained local ML model, every client calculates the local logit representing the inferred probability that each data sample is classified into each class.
Subsequently, each local logit is aggregated to each client on a per-label basis in the ``2.~Prediction \& Logit aggregation per label'' step.
The aggregated logits are uploaded to a remote server in the ``3.~Upload'' step.
The uploaded logits are aggregated in the server in the ``4.~Aggregation'' step, whereas the aggregated logits are broadcast to the clients in the ``5.~Broadcast'' step.
Finally, in the ``6.~Distillation'' step, each client re-trains its local model using both the pre-attached labels and the broadcasted logits.

The detailed FD procedures are as follows.
In the following, we consider that $k$th client's private dataset can be divided into $N_L$ set $(\bm{D}_{k,n_l})_{n_l=1}^{N_L}$,
where the sample pertaining to the class $n_l$ is categorized to set $\bm{D}_{k,n_l}$, and
$N_L$ is the number of objective classes.

\vspace{.3em}\noindent\textbf{2.~Prediction \& Logit aggregation per label.}\quad
Each client $k$ predicts logit of its data input and calculate local-average logit $\bm{t}_{k,{n_l}}$ for each class, as follows:
\begin{align}
	\bm{t}_{k,n_l} = \frac{1}{|\bm{D}_{k,n_l}|} \sum_{\{\bm{d},\bm{t}\} \in \bm{D}_{k,n_l} } {F(\bm{d}\,|\,\bm{w}_k)}. 
\end{align}
If the client dose not have any sample pertaining to class $n_l$, $\bm{t}_{k,n_l} = \bm{0}$.

\vspace{.3em}\noindent\textbf{3.~Upload, 4.~Aggregate, and 5.~Broadcast.}\quad
In the following ``3.~Upload'' step, the local-average logit $\bm{t}_{k,{n_l}}$ are uploaded from each client to the server.
In the ``4.~Aggregate'' step, the uploaded logits are aggregated to create the global-average logit $\bm{t}_{\mathrm{g},n_l}$ per class as follows:
\begin{align}
	\bm{t}_{\mathrm{g},{n_l}} = \frac{1}{|\bm{K}_{n_l}|} \sum_{k \in \bm{K}_{n_l}}  \bm{t}_{k,n_l},
\end{align}
where $\bm{K}_{n_l}$ is the subset of clients having any sample pertaining to class $n_l$.
Subsequently, the server broadcasts the global-average logit to all the clients via multicast channels in the ``5.~Broadcast'' step.

\vspace{.3em}\noindent\textbf{6.~Distillation.}\quad
Each client $k$ updates the model parameters using the pre-attached labels and the distillation logits: $\bm{T}_k$ and $\hat{\bm{T}}_k$,
where $\hat{\bm{T}}_k = \{\hat{\bm{t}}_{k,i}\}_{i=0}^{I_k}$.
The distillation logit $\hat{\bm{T}}_k$ is obtained using the broadcasted global-average logit and the local-average logit.
In more detail, considering that the sample $\bm{d}^{\mathrm{p}}_{k,i}$ pertains to the class $n_l$, $\hat{\bm{t}}_{k,i}$ is obtained as follows:
\begin{align}
	\hat{\bm{t}}_{k,i} =\frac{1}{|\bm{K}_{n_l}|-1} ( |\bm{K}_{n_l}|\,\bm{t}_{\mathrm{g},n_l} - \bm{t}_{k,n_l}).
\end{align}
Using $\bm{T}_k$ and $\hat{\bm{T}}_k$, the model is updated as follows:
\begin{align}
	\!\!\!\!\!\!\!\!\!\! \bm{w}_k \gets \bm{w}_k - \eta \nabla \{\phi (\bm{D}_k^{\mathrm{p}}, \bm{T}_k\,|\,\bm{w}_k) + \gamma \phi (\bm{D}_k^{\mathrm{p}}, \hat{\bm{T}}_k\,|\,\bm{w}_k) \},
\end{align}
where $\gamma$ is a weight parameter for the distillation regularizer.
All the procedures, except for ``1.~Update,'' are iterated for a finite number of rounds.

\vspace{.3em}\noindent\textbf{Negative effect of Federated distillation under non-IID data.}\quad
\begin{figure}[t!]
	\centering
	\includegraphics[width=0.35\textwidth]{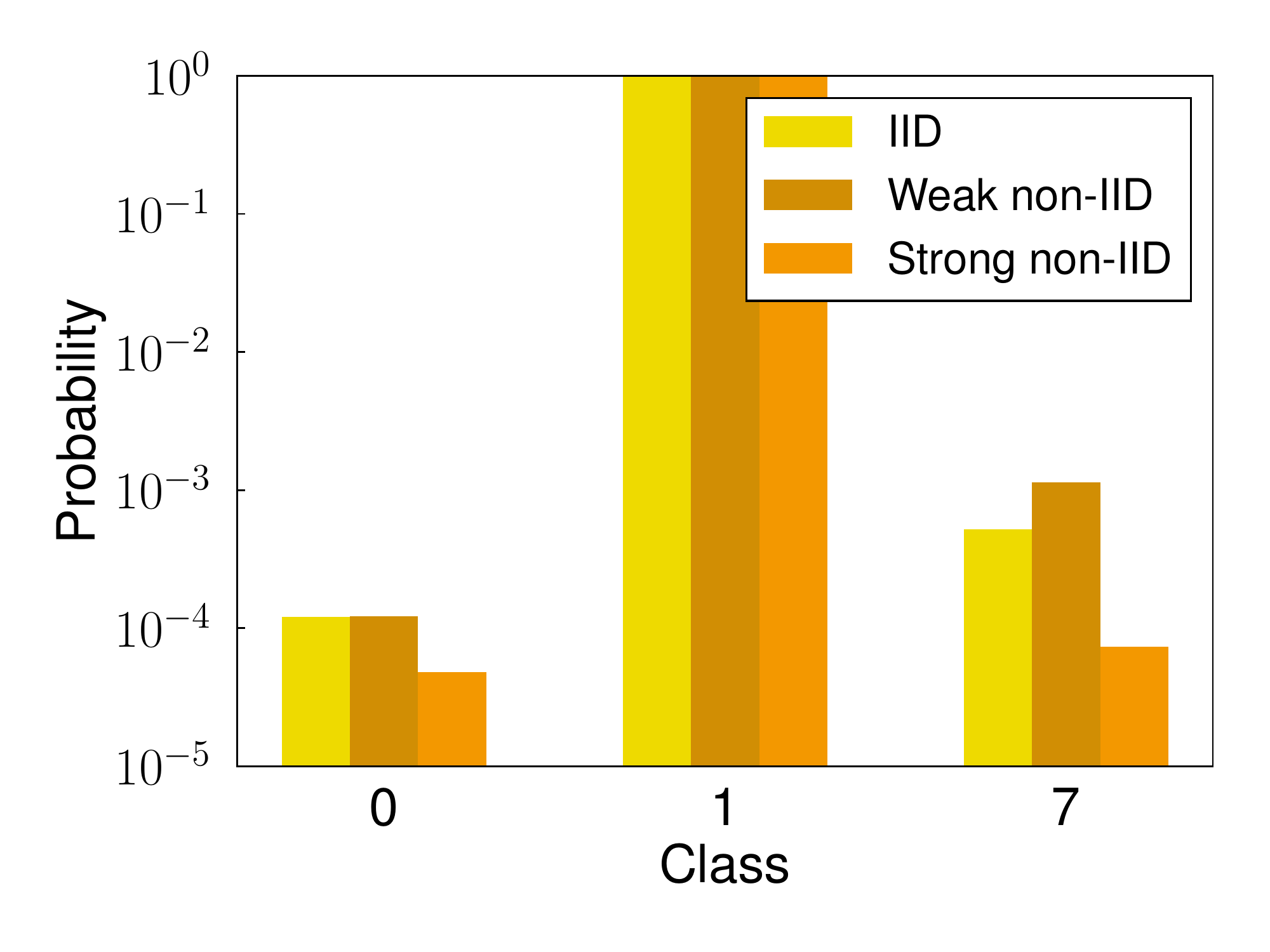}
	\caption{A part of FD's global logit after 16 rounds, using MNIST dataset, under IID, weak non-IID, and strong non-IID data distribution. That is, probabilities of class ``0,’’ ``1'' and ``7'' of the logit for class ``1.''}
	\label{fig:FD_global_logit}
\end{figure} 

The model performance trained by FD is much lower than FL under non-IID data.
The reason is that the global logits of FD retain similar information to local labels already attached to each mobile device.
The global logit is calculated as the average of the local per-label average logits among the clients.
In the following, we explain the method to obtain global logit in FD more formally.
We consider that $k$th client's private dataset can be divided into $N_L$ subset $(\bm{D}_{k,n_l})_{n_l=1}^{N_L}$, according to the ground-truth labels, where $N_L$ is the number of objective classes.
The global logit for class $n_l$ can be represented as follows:
\begin{align}
	\bm{t}_{\mathrm{g},n_l} = \frac{1}{|\bm{K}_{n_l}|} \sum_{k \in \bm{K}_{n_l}} \frac{1}{|\bm{D}_{k,n_l}|} \sum_{\{\bm{d},\bm{t}\} \in \bm{D}_{k,n_l} } {F(\bm{d}\,|\,\bm{w}_k)},
\end{align}
where $\bm{K}_{n_l}$ is the subset of clients having any class $n_l$ sample, and $\bm{w}_k$ is the model weights of the $k$th client.
Considering the strong non-IID data, where each client only has one or two class sample,
$F(\bm{d}\,|\,\bm{w}_k)$ is almost the same as the one-hot label $\bm{t}$, which is the ground-truth label already attached to $\bm{d}$.
This is because of the over-fitting of $\bm{t}$, due to the use of few samples.
Thus, $\bm{t}_{\mathrm{g},n_l}$, which is the average of the almost one-hot logit $F(\bm{d}\,|\,\bm{w})$, is also similar to the one-hot labels.

To support the statement, we analyzed FD's global logits under three data distributions; IID, weak non-IID, and strong non-IID.
The number of clients $K$ was fixed to ten.
The strong non-IID distribution used in our original manuscript implies that each client has a dataset consisting of two or three classes' samples.
The week non-IID distribution implies that the clients have the datasets consisting of ten classes, and the number of data samples on a few of the classes is much smaller than that of the other classes.

Fig.~\ref{fig:FD_global_logit} shows a part of the FD's global logit under three data distribution using MNIST, 
i.e., class probabilities of class ``0,'' ``1,'' and ``7'' of the global logit for class ``1.''
Obviously, the largest probability of the logit is class ``1.''
Under strong non-IID data, the probability of class ``7'' is as large as that of ``0'' and smaller than that under IID or weak non-IID.
This implies that the global logit under strong non-IID is more similar to one-hot than that of IID or week non-IID.
This may be the reason that the FD model performance is much lower than that of FL under strong non-IID data.

Noted that, under IID or weak non-IID data, the probability of class ``7'' is large than that of ``0,'' which results in the success of the FD.
This difference between class ``0'' and class ``7'' indicates that the digit ``1'' image is more similar to the digit ``7'' than the digit ``0,'' which is essential to the success of the knowledge distillation as supported in~\cite{distillation}.
Hence, we can conclude that the intensity of non-IID data distributions exactly affects the model performances.

\subsection{Proposed Distillation-Based Semi-Supervised Federated Learning}
\label{ssec:Overview}

\subsubsection{Background and Overview of Distillation-Based Semi-Supervised Federated Learning}
\label{sssec:background}
The proposed DS-FL is motivated by the lower performance of the model trained following the FD benchmark presented in the previous section.
In FD, the global logits are used to distinguish the class to which each sample in the local dataset belongs.
However, as the dataset is already labeled, the ``6.~Distillation'' step can result in a similar model to the one trained in the previous ``1.~Update'' step (see Fig.~\ref{fig:DS-FL_process}~(b) and compare ``1.~Update'' and ``6.~Distillation'').
Hence, the models trained in FD exhibit similar performance to the local model training, which is lower than the benchmark~1, i.e., FL, with model parameter exchange.

Hence, the fundamental idea behind the proposed DS-FL is to share the unlabeled dataset and use global logit to identify what class each sample in the unlabeled dataset pertains.
This creates a new labeled dataset, based on which the local ML model is further trained, as shown in ``6.~Distillation'' step in Fig.~\ref{fig:DS-FL_process}~(c).
Due to this training procedure, the proposed DS-FL avoids the similarly in training between ``1.~Update'' and ``6.~Distillation'' steps can enhance the model performance benefitting from data augmentation effects.
In other words, DS-FL uses unlabeled distillation, where a neural network model is trained using other models' prediction of the unlabeled data.
In the following, we detail the training procedure of the proposed DS-FL.

\subsubsection{Detailed Procedure of Distillation-Based Semi-Supervised Federated Learning}
\label{sssec:procedure_dsfl}

The detailed procedure of DS-FL is depicted in Fig.~\ref{fig:DS-FL_process}~(c).
In the following, we consider that each client $k = 1, 2, \dots, K$ does not only hold the labeled private dataset $(\bm{d}^{\mathrm{p}}_{i, k}, \bm{t}_{i, k})_{i = 1}^{I_k}$, but also the shared unlabeled dataset $(\bm{d}^\mathrm{o}_j)_{j = 1}^{I^{\mathrm{o}}}$, where $\bm{d}^{\mathrm{p}}_{i, k}$ and $\bm{d}^\mathrm{o}_j$ denote the vectorized input samples in the labeled and unlabeled datasets, respectively.
Moreover, $I_k$ and $I^{\mathrm{o}}$ denote the number of samples in the labeled and unlabeled datasets, respectively.
Additionally, $\bm{o}_r \subset \{1,2,\dots, I^\mathrm{o}\}$ represents index set of the unlabeled dataset, where $r$ indicates round index.
Moreover, $I^{\mathrm{o_r}}$ represents the size of $\bm{o}_r$.
Considering $N_\mathrm{L}$ as the number of objective class,
the term $\bm{t}_{i, k} = [t_{i, k, 1}, \dots, t_{i, k, N_{\mathrm{L}}}]^{\mathrm{T}}$ is the vectorized form of the label attached to the sample $\bm{d}^{\mathrm{p}}_{i, k}$ and is in the one-hot representation, wherein the element $t_{i, k, n}$ equals 1 if the $n$th label is the ground-truth and 0 otherwise.
For shorthand notation, let $N_{\mathrm{S}}\times I_k$  and $N_{\mathrm{S}}\times I^{\mathrm{o}}$ matrices $\bm{D}_k^{\mathrm{p}}$ and $\bm{D}^{\mathrm{o}}$ denote the concatenations of $(\bm{d}^{\mathrm{p}}_{i, k})_{i = 1}^{I_k}$ and $(\bm{d}^{\mathrm{o}}_{j})_{j = 1}^{I^{\mathrm{o}}}$, respectively, and  let $N_{\mathrm{L}}\times I_k$ matrix $\bm{T}_k$ denote the concatenation of $(\bm{t}_{i, k})_{i = 1}^{I_k}$, where $N_{\mathrm{S}}$ denote the dimension of the input samples.
Additionally, let $N_{\mathrm{S}}\times I^{\mathrm{o_r}}$ matrix $\mathbf{D}^{\mathrm{o}_r}$ denotes the subset of the unlabeled dataset $(\mathbf{d}^{\mathrm{o}}_j)_{j \in \mathbf{o}_r}$.
The index set $\bm{o}_r$ is determined randomly by the server and shared among the clients before ``2.~Prediction'' step.

\vspace{.3em}\noindent\textbf{1.~Update.}\quad 
In the ``1.~Update'' step, each client updates its model with its private dataset, as shown in~\eqref{equ:model_update}.

\vspace{.3em}\noindent\textbf{2.~Prediction.}\quad 
Based on the model learned in the previous step, each client predicts the local logit, i.e., the labels for data samples in a shared unlabeled dataset.
More specifically, given the model parameter $\bm{w}_k$, and $\mathbf{o}_r$, each client predicts local logits $\hat{\bm{t}}_{j, k}$ for $j \in \mathbf{o}_r$ as follows:
\begin{align}
    \label{equ:local_logit_calclation}
    \hat{\bm{t}}_{j, k} = F(\bm{d}^\mathrm{o}_j\,|\bm{w}_k\,).
\end{align}
For shorthand notation, the $N_{\mathrm{L}}\times I^{\mathrm{o_r}}$ matrix $\hat{\bm{T}}_k$ denotes the concatenation of $(\hat{\bm{t}}_{i, k})_{i\in \bm{o}_r}$.

\vspace{.3em}\noindent\textbf{3.~Upload.}\quad 
The local logits $\hat{\bm{T}}_k$ are uploaded from each client to the server, differing from FL with model parameter exchange that uploads the model parameters $\hat{\bm{w}}_k$.

\vspace{.3em}\noindent\textbf{4.~Aggregation and 5.~Broadcast.}\quad 
The server aggregates the logits from clients to create global logits $\hat{\bm{T}}$.
The procedure for aggregating uploaded logits is described in Section~\ref{sec:make_pseudo_label}.
Subsequently, the server broadcasts global logits to all the clients via multicast channels.

\vspace{.3em}\noindent\textbf{6.~Distillation.}\quad 
The clients update their local model based on the broadcasted global logits $\hat{\bm{T}}$ and shared unlabeled dataset $\bm{D}^{\mathrm{o}_r}$.
More concretely, the model parameters are updated as follows:
\begin{align}
    \label{equ:model_distil}
    \bm{w}_k \gets \bm{w}_k - \eta_{\mathrm{dist}} \nabla \phi (\bm{D}^{\mathrm{o}_r}, \hat{\bm{T}}\,|\,\bm{w}_k),
\end{align}
where $\eta_{\mathrm{dist}}$ is the learning rate in the proposed distillation procedure.
In addition to the clients' local models, the server has a global model $\bm{w}_\mathrm{g}$.
The server updates the global model based on the broadcasted global logits $\hat{\bm{T}}$ and shared unlabeled dataset $\bm{D}^{\mathrm{o}_r}$.
More concretely, the model parameters are updated as follows:
\begin{align}
	\label{equ:model_distil_server}
	\bm{w}_\mathrm{g} \gets \bm{w}_\mathrm{g} - \eta_{\mathrm{dist}} \nabla \phi (\bm{D}^{\mathrm{o}_r}, \hat{\bm{T}}\,|\,\bm{w}_\mathrm{g}).
\end{align}
In Section 4, the global model is used for evaluating the performance of the DS-FL framework.

These procedures are iterated for a finite number of rounds.
The overall procedures are summarized in Algorithm~\ref{alg:DS-FL}.

\begin{algorithm}[t]
    \caption{DS-FL}         
    \label{alg:DS-FL}         
    \begin{algorithmic}[]
    \\
	\Initialization
    \State Initialize all the client models $\bm{w}_k$ and the global model $\bm{w}_\mathrm{g}$
    \State Distribute the open data $\bm{D}^\mathrm{o}$ to all clients
    
    \ClientUpdate
    \For {Each client $k$ \textbf{in parallel}}
    \State Update the local model parameter $\bm{w}_k$ via \eqref{equ:model_update}
    \EndFor
    
    \ClientPrediction
    \State All the clients share the index set $\mathbf{o}_r$
    \For {Each client $k$ in parallel}
    \State Calculate local logits $\hat{\bm{T}}_{k}$ via \eqref{equ:local_logit_calclation}
    \EndFor
    
    \ClientUpload
    \State Each client uploads the local logits $\hat{\bm{T}}_k$
    
    \Aggregation 
    \State Server aggregates the logits to create the global logit $\hat{\bm{T}}$
    \State according to \eqref{equ:ERA} in ERA (proposed)
    \State or \eqref{equ:SA} in SA (baseline)
    
    \Broadcast
    \State Broadcast $\hat{\bm{T}}$ to all clients
    
    \Distillation
    \For {Each client $k$ \textbf{in parallel}}
    \State Update the local model parameter $\bm{w}_k$ via \eqref{equ:model_distil}
    \EndFor
    
    \Itaration
    
    \end{algorithmic}
\end{algorithm}

\section{Entropy Reduction Aggregation}
\label{sec:make_pseudo_label}

This section presents the proposed logit aggregation method, i.e., ERA, which intentionally reduces the entropy of global logits.
We define the entropy of a logit $\bm{t}$ as follows:
\begin{align}
    f_\mathrm{e}(\bm{t}) = -\sum_{n=1}^{N_\mathrm{L}} t_n \log  {t_n}.
\end{align}
First, we detail our motivation to propose ERA by highlighting that the simple baseline-aggregation-method, involving averaging only the local logits, results in a higher entropy of global logits in the heterogeneity of data distributions among clients than those without such heterogeneity.
Subsequently, we detail how to reduce entropy in ERA.

\subsection{Motivation for Entropy Reduction Aggregation}
\label{ssec:motivation_era}
\begin{figure}[t]
    \centering
    \includegraphics[width=0.4\textwidth]{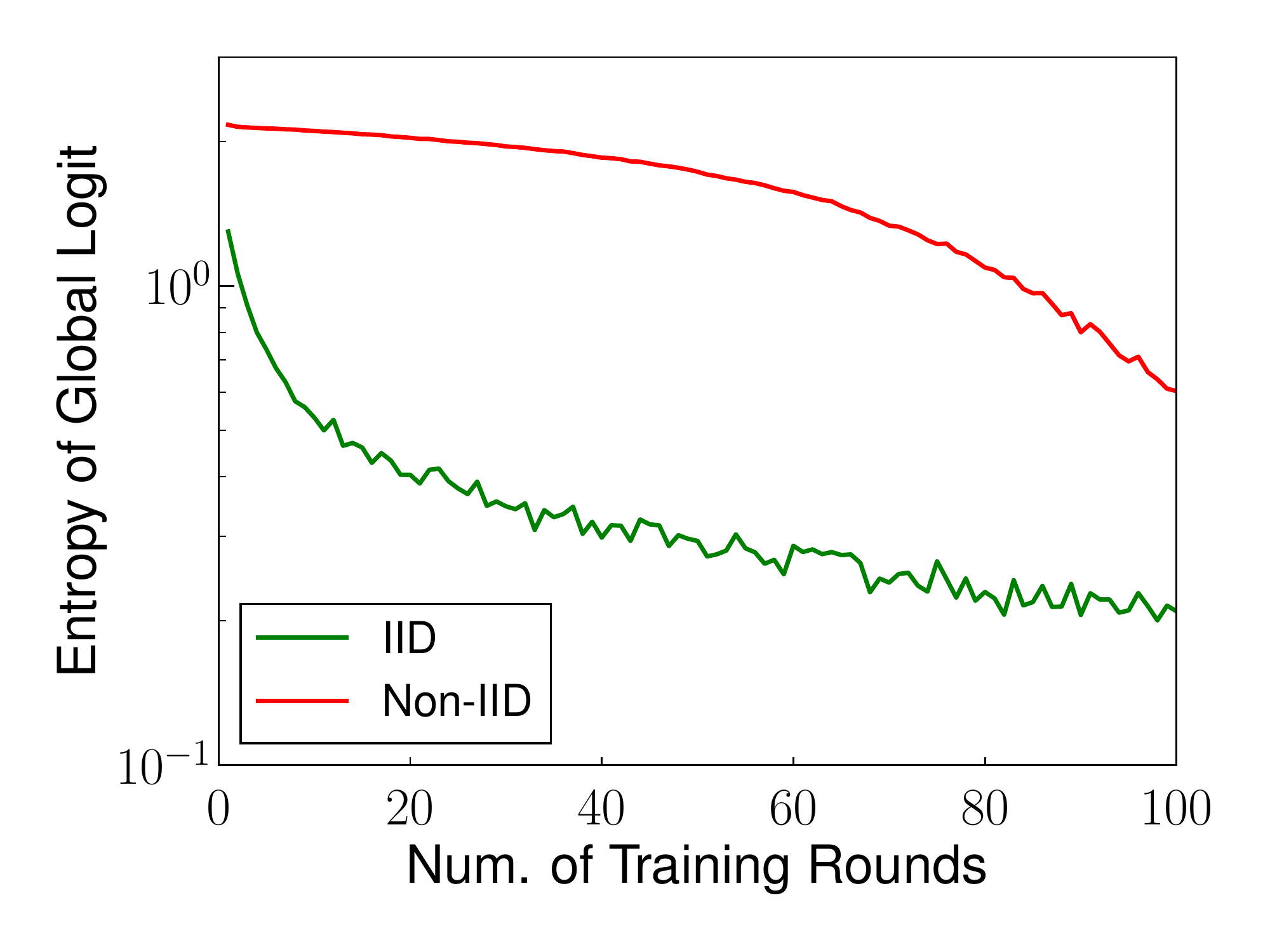}
    \caption{Entropy of the global logit in SA method vs. training rounds. For IID and non-IID data, using MNIST dataset.
    }
    \label{fig:EA_IID_nonIID}
\end{figure}

The motivation for reducing the entropy of global logits is to accelerate and stabilize DS-FL, particularly in non-IID data distributions.
In the collaborative learning with non-IID data, the entropy of global logits is much larger than appropriate ones.
Fig.~\ref{fig:EA_IID_nonIID} shows the comparison of global logits yielded via the simple aggregation method involving the averaging of the uploaded local logits for IID and non-IID data distributions.
Under non-IID data, the entropy in the early stage of training is higher than 2.0, which is approximately the upper limit of the entropy in the ten-class classification problem.
These maximum values of the entropy are meaningless because they do not identify to what class each input sample pertains.
Hence, it is difficult to train using such inappropriately high entropy logit in the simple aggregation method, and hence, the reduction of the entropy is required for training success.

Another favorable consequence of reducing the entropy of global logits is to enhance the robustness against various attacks corrupting local logits and noising open data.
In DS-FL, a malicious client can upload local logits that do not enhance or even harm model performance, which can occur by, for example, updating the local models over the dataset that is not labeled properly.
In addition to such noisy label, open data can be noised; for example, the inadequate data is added to the open data.
In these cases, similarly to under non-IID data distributions, the entropy of global logits yielded from the simple aggregation method averaging local logits becomes higher, leading to poorer model performances.
This fact is verified in Section~\ref{sec:experiment}.
Hence, reducing the entropy of the global logit is expected to enhance the robustness against such attacks.

\subsection{Procedure for Reducing Entropy}

\begin{figure}[t]
    \centering
    \subfloat[Procedure of proposed ERA.]{\includegraphics[width=0.35\textwidth]{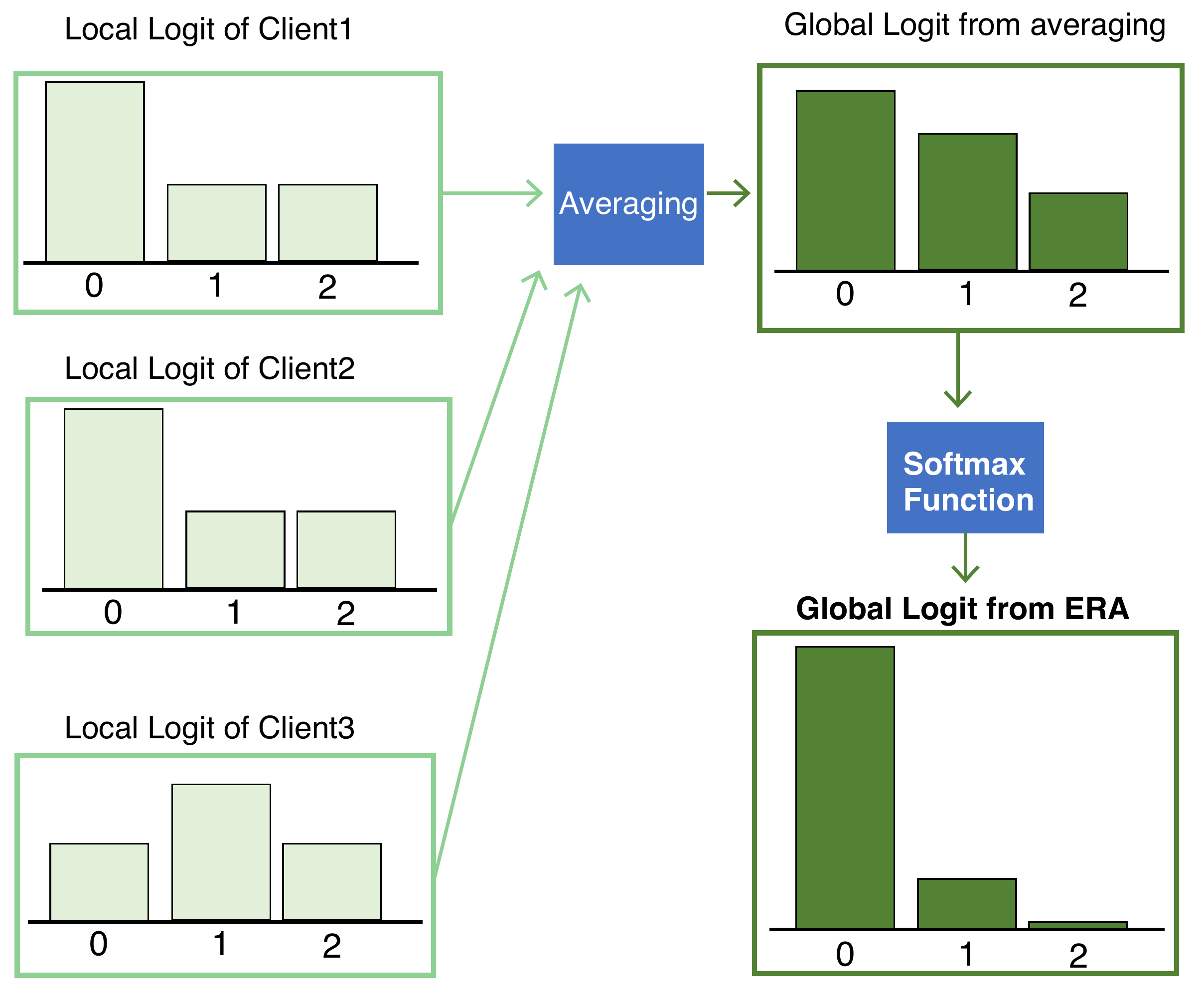}}\\
    \subfloat[Entropy comparison between the input and output of the softmax function.]{\includegraphics[width=0.3\textwidth]{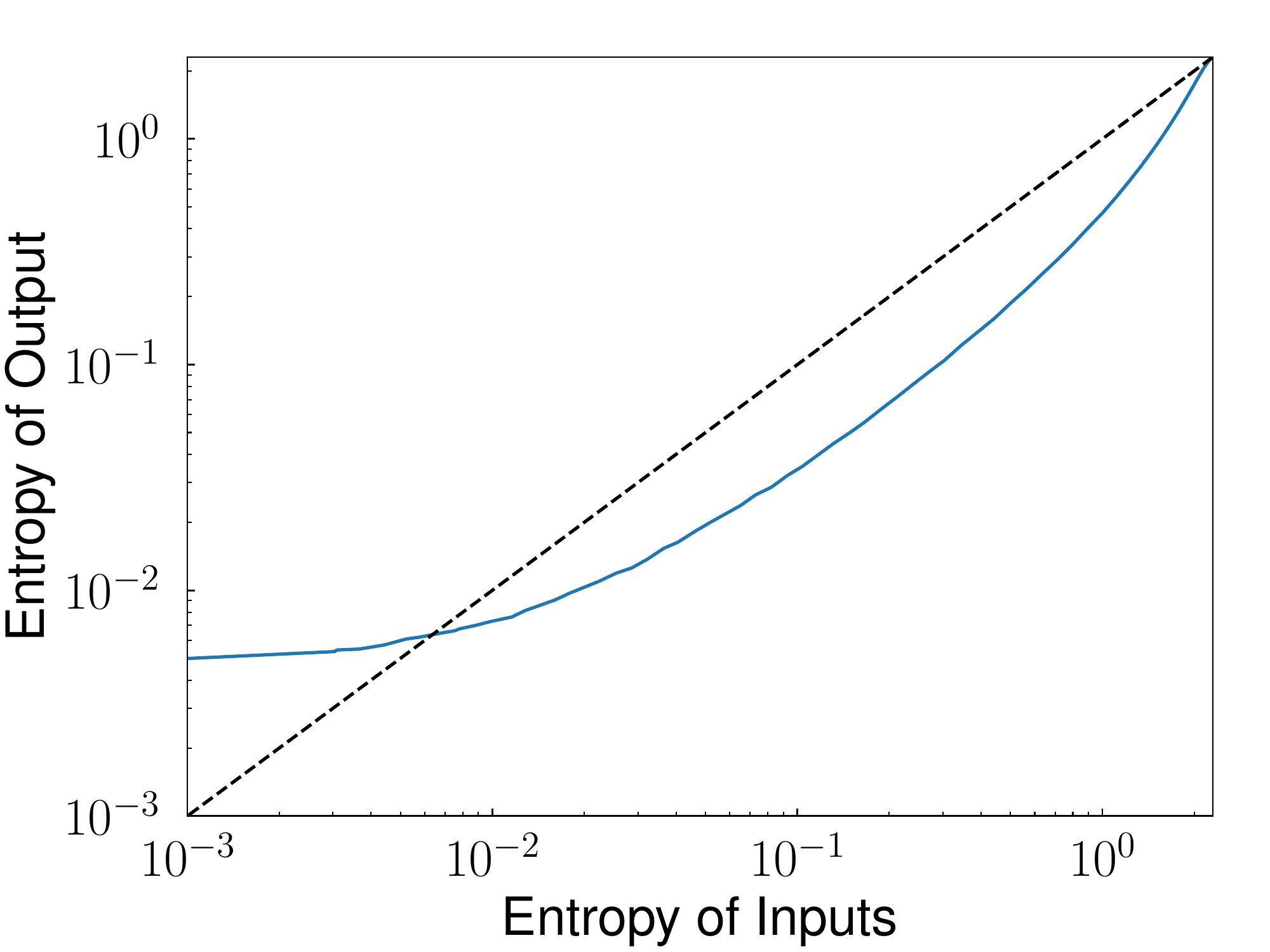}}
    \caption{Illustrative example of proposed ERA at $T = 0.1$ on three-class classification task.}
    \label{fig:GL_logit_Gen}
\end{figure}

The proposed ERA reduces the entropy of the global logits yielded from averaging uploaded local logits via the procedure depicted in Fig.~\ref{fig:GL_logit_Gen}~(a).
To reduce the entropy of the global logit, we use the softmax function as an example.
Let $\hat{\bm{T}}^{(\mathrm{ERA})}$ denote the global logit yielded from ERA, which is an $N_{\mathrm{L}}\times I^{\mathrm{o_r}}$ matrix.
Given the temperature of the softmax function $T$, the global logit generated by ERA is described as follows:
\begin{align}
    \label{equ:ERA}
    \hat{\bm{T}}^{(\mathrm{ERA})} = F_\mathrm{s}\!\left(\frac{1}{K} \sum_{k=1}^{K} \hat{\bm{T}}_k \,\middle|\, T\right),
\end{align}
where $F_{\mathrm{s}}(\cdot\,|\, T)$ denotes the softmax function with respect to temperature $T$.
The softmax function $f_{\mathrm{s}}(\bm{t}\,|\, T) : \mathbb{R}^{N_\mathrm{L}} \rightarrow \mathbb{R}^{N_\mathrm{L}}$ is denoted as follows:
\begin{align}
    f_{\mathrm{s}}(\bm{t}\,|\, T) =  \frac{1}{\sum_{n=1}^{N_\mathrm{L}} \mathrm{e}^{\frac{t_n}{T}}}  \mathrm{e}^\frac{\bm{t}}{T}.
\end{align}
Moreover, $F_{\mathrm{s}}(\bm{T}\,|\, T)$ is denoted as follows:
\begin{align}
    F_\mathrm{s}(\bm{T}\,|\, T) 
    = \{f_{\mathrm{s}}(\bm{t}_1\,|\, T),\cdots,f_{\mathrm{s}}(\bm{t}_{I_r}\,|\, T) \}.
\end{align}
The higher the temperature, the higher the entropy of the output of the softmax function, and viceversa.

For example, ERA sets lower temperature of $T=0.1$,
while original knowledge distillation (KD)~\cite{distillation} sets a higher temperature of $T = 20$.
Hinton et al.~\cite{distillation} state that in KD, the scores in the logits can be interpreted as an inherent similarity between the corresponding label and the input samples.
For example, one handwritten digit labeled as ``7'' might provide a score of $0.1$ to the label ``1'' and that of $0.01$ to the label ``4.''
Given that the physical meaning of this score is the probability that each sample pertains to the corresponding label, these scores can be interpreted as how much the given handwritten digit is prone to be ``1'' and ``4.''
As the above bias of this likeliness (0.1 to ``1'' and 0.01 to ``4'') may be yielded due to the similarity between ``7'' and ``1,'' the given logit scores are interpreted as the similarity between the corresponding label and the input samples.
To transfer a similar structure,
we trained the student model using the logit create by the softmax function using a high temperature; thus,
emphasizing the non-highest scores.
In contrast, particularly under non-IID data, the global logits of DS-FL is much more ambiguous than the predictions of a well-trained teacher model such as the model used in KD~\cite{distillation}.
Moreover, the ambiguity may conceal the true classes each sample pertains, wherein the lower temperature is expected to be useful to balance the class information and a similar structure.
As shown in Fig.~\ref{fig:GL_logit_Gen}~(b), at the temperature of $T =0.1$, the entropy values outputted by the softmax function are generally lower than the input.
Hence, as shown in Fig.~\ref{fig:GL_logit_Gen}~(a), the global logit yielded from ERA is sharper than that yielded via averaging the uploaded local logits.
Noted that $T$ of ERA discussed here differs from the temperature of the softmax function that activates the fully-connected layer on the output side of neural networks.
The former temperature is specific to the training procedure and is set to a certain value lower than 1.0 (e.g. 0.1 in the experimental evaluation), 
whereas the latter temperature is set to 1.0 in the training and inference steps.

The ERA is compared with the baseline of only averaging uploaded local logit, which is named simple aggregation (SA).
The resultant global logit yielded from SA $\hat{\bm{T}}^{(\mathrm{SA})}$ is given as follows.
\begin{align}
    \label{equ:SA}
    \hat{\bm{T}}^{(\mathrm{SA})} = \frac{1}{K} \sum_{k=1}^{K} \hat{\bm{T}}_k.
\end{align}

\section{Experimental Evaluation}
\label{sec:experiment}

\subsection{Setup}
\label{ssec:ML_Task}
\vspace{.3em}\noindent\textbf{Datasets.}\quad 
Four datasets were used for evaluation purposes, including image classification and text classification.
For image classification, two major tasks, MNIST~\cite{MNIST} and Fashion-MNIST~\cite{F_MNIST}, were used.
MNIST~\cite{MNIST} is a widely-used object classification dataset consisting of $60{,}000$ training images and $10{,}000$ testing images with 10 image classes.
Fashion-MNIST~\cite{F_MNIST} comprises $60{,}000$ training images and $10{,}000$ testing images of 10 different fashion products such as coats and sneakers.
These datasets have been used in several machine learning studies.
In addition to the image classification tasks, two major text classification tasks, Internet movie database (IMDb) review-sentiment and Reuters datasets, are used for evaluation purposes.
The IMDb dataset was created in~\cite{maas2011learning}, which consists of $50{,}000$ textual reviews of movies and divided into $25{,}000$ training dataset and $25{,}000$ testing dataset.
The reviews are categorized into negative and positive sentiment class.
The Reuters dataset
\footnote{The Reuters dataset we used was Keras~\cite{chollet2015keras} revised subset of Reuters-21578 corpus~\cite{david2004reuters}, \cite{david2004reuters} is freely available for experimentation purposes from \url{http://www.daviddlewis.com/resources/testcollections/~reuters21578/}.} consists of $11{,}228$ headline articles, divided into training and testing dataset in a ratio of 8:2.
The headlines are categorized into 46 classes of topics such as ``earn'' and ``trade.''

\vspace{.3em}\noindent\textbf{Pre-processing of the sentences.}\quad
We used different neural network architecture and different preprocessing method for IMDb and Reuters datasets.
For the IMDb dataset, we only considered the top 20k words and the first 200 words of each movie review, following Keras~\cite{chollet2015keras} tutorial.
According to the frequency appearance, each word is converted to an integer.
Moreover, following the word order in the sentences, each sentence is converted to a sequence of integers.
For Reuters dataset, we only considered the top 10k words and each word was converted to a integer, as IMDb dataset.
In the sentences categorization task, such as Reuters dataset, the type of words are more useful than the word order.
Thus, we employed the Bag-of-Words method, which is often used as a preprocessing for the Reuters dataset~\cite{zhao2018fuzzy},
i.e., the headline is converted to a binary vector, indicating the sentence composition of words.

\vspace{.3em}\noindent\textbf{Data partitions.}\quad 
The data distribution over the clients was determined based on~\cite{FedAve}.
For the image classification tasks, we fixed the number of clients $K$ to 100.
Subsequently, the dataset was divided into the unlabeled open dataset and the labeled private dataset.
Let denote that the open dataset consists of $I^\mathrm{o}$ images, and the private dataset consists of $I^\mathrm{p}$ image-label pairs,
where $I^\mathrm{o} + I^\mathrm{p} \leq 60{,}000$.
This study considered two ways of partitioning the private dataset over the clients: IID datasets and non-IID datasets.
The private dataset was shuffled and partitioned into $K$ portions for the clients to obtain the IID datasets.
Thus, each client had $I^\mathrm{p}/{K}$ pairs of images and labels.
To generate non-IID datasets, we sorted the private dataset by its classification label and divided it into $2K$ shards of size $I^\mathrm{p}/{2K}$, among which two shards are assigned to each client.
In this study, the non-IID data distribution among the clients followed the pioneer study~\cite{FedAve}, which considered more severe non-IID data than that in~\cite{FD}.
Hence, the differences in the data distribution are the reason for the difference in the test accuracy of the FD between this study and~\cite{FD}.
In the evaluation, we did not adapt any data augmentation, such as rotation or flipping of an image.

For the text classification task, we fixed the number of clients $K$ to 10.
Subsequently, the dataset was divided into the unlabeled open dataset and the labeled private dataset.
For IMDb dataset, $(I^\mathrm{o}, I^\mathrm{p})$ was $(10{,}000, 15{,}000)$ and for Reuters dataset, $(I^\mathrm{o}, I^\mathrm{p})$ was $(3{,}982, 5{,}000)$.
To generate the non-IID partitioned datasets for IMDb, we divided the dataset so that,
for all clients, the ratio of the number of positive labeled sentences to that of negative was 9:1 or 1:9.
In consequence, some clients had 150 positive labeled sentences and $1{,}350$ negative labeled sentences,
and the other had $1{,}350$ positive labeled sentences and 150 negative labeled sentences.
To generate non-IID partitioned datasets for Reuters,
we sorted the private dataset by its classification label and divided it into $K$ shards of size $I^\mathrm{p}/{K}$, among which one shard was assigned to each client.

Our evaluation aimed to evaluate FL frameworks under the same clients' privacy level.
Thus, we avoided to compare with other baselines sharing any clients' labeled dataset, such as FD+FAug~\cite{FD}.

\vspace{.3em}\noindent\textbf{ML model.}\quad 
For the image classification tasks, we examined two ML models designed for either MNIST or Fashion-MNIST dataset.
Specifically, the model for MNIST was a convolutional neural network model that consisted of two $5 \times 5$ convolution layers
(32 and 64 output channels, each of which was activated by batch normalization and ReLU, followed by $2 \times 2$ max pooling) and two fully-connected layers (512 units with ReLU activation and another 10 units activated by softmax).
For Fashion-MNIST, the model consisted of six $3 \times 3$ convolution layers 
(32, 32, 64, 64, 128, and 128 channels, each of them activated by ReLU and batch normalized. Every two of them followed by $2 \times 2$ max pooling) and by three entirely connected layers (382 and 192 units with ReLU activation and another 10 units activated by softmax).
For the text classification tasks, we examined two ML models designed for either IMDb or Reuters dataset.
The IMDb dataset is semantic classification, where the word order in the sentences is useful.
Thus, for the IMDb dataset, we employed long short-term memory (LSTM),
which can learn the time-dependent relationships among inputs and outputs.
Specifically, the model for IMDb consisted of a simple LSTM model, which followed Keras~\cite{chollet2015keras} tutorial, consisting of a embedding layer (output dimension of each word was 32), a LSTM layer (32 nodes), and a fully-connected layer (2 unit was activated by softmax).
For Reuters dataset, we employed a simple multi-layer perceptron (text-DNN).
Specifically, the model was a three layer perceptron (512 and 128 units activated by ReLU and batch normalized and another 46 units activated by softmax).
The ML models resulted in $583{,}242$ model parameters (2.3 megabytes in a 32-bit float) for the MNIST dataset, $2{,}760{,}228$ model parameters (11.2 megabytes in a 32-bit float) for the Fashion-MNIST dataset, $646{,}338$ model parameters (2.6 megabytes in a 32-bit float) for the IMDb dataset, and $5{,}194{,}670$ (20.8 megabytes in a 32-bit float) model parameters for Reuters dataset.

\vspace{.3em}\noindent\textbf{Training hyperparameters.}\quad 
When the models were updated and distilled, the optimizer, mini-batch size, the number of epochs in each round, and training rate were selected as stochastic gradient decent, 100, 5, and 0.1, respectively.
For text classification tasks, they are selected as Adam, 128, 5, and 0.001, respectively.
The temperature of the softmax function $T$ in the DS-FL with ERA was set to 0.1.
As shown in Section 3.2, the temperature of the softmax function that activates the fully-connected layer on the output side of neural networks is set to 1.0 in the training and inference steps.
The amount of unlabeled data used in each round, i.e., the size of $\mathbf{o}_r$, was $1{,}000$. 

\vspace{.3em}\noindent\textbf{(2-7) Attack settings.}\quad 
To evaluate the attack robustness of ERA, we considered attacks where malicious clients corrupted local logits.
Specifically, we considered noisy labels, noisy data, and model poisoning attacks.
The robustness evaluations of this section used image classification task described above.

\vspace{.3em}\noindent\textbf{Noisy labels.}\quad 
First, in the noisy label attack, a particular client's images pertaining to a certain class were labeled as another class to corrupt the local logit.
We assumed that all clients had the same degree of noisy labeled datasets, i.e., mistakenly labeled private data. 
This was regarded as a situation where all clients can be considered as attackers.
Thus, we evaluated DS-FL and FL in a worst-case scenario, which we believed was sufficiently worthwhile to understand the attack robustness in practical and severe situations.
More precisely, consider a number of noising class $C$, where each client independently selects $C$ classes as source classes $\bm{S} = \{S_1,\dots,S_{C}\}$ and another $C$ classes as false classes $\bm{F} = \{F_1,\dots,F_{C}\}$. Subsequently, all the images pertaining to the source class $S_c \in \bm{S}$ are mistakenly classified to the corresponding false class $F_c \in \bm{F}$.
Assuming 10 objective classes, $CI^\mathrm{p}/{10}$ images of the private dataset were mistakenly labeled.
In this evaluation, we used MNIST dataset, considering IID data distribution.

\vspace{.3em}\noindent\textbf{Noisy data.}\quad 
Second, we evaluated the robustness of ERA in a noisy data attack, where a malicious client adds noisy semantic data into the open dataset.
In more detail, consider training a handwritten digit classifier, where the private datasets and the test dataset, testing our methods, are the MNIST dataset.
In this experiment, we assumed non-IID distribution and
 added $I^\mathrm{n}$ Fashion-MNIST images to $I^\mathrm{o}$ MNIST open dataset,
i.e., $I^\mathrm{o} + I^\mathrm{n}$ images were used for the unlabeled open dataset.
We fixed $I^\mathrm{o}$ to $20{,}000$ and experimented with $I^\mathrm{n}$.

\vspace{.3em}\noindent\textbf{Model poisoning.}\quad
To evaluate robustness against model poisoning attack, we conducted the experiment with a malicious client and 99 benignant clients.
In this analysis, we assumed that the malicious client performed a model poisoning attack~\cite{backdoorFL}, which aimed to replace the global model with an arbitrarily model and introduced a backdoor to the global model.
In this attack, the malicious client aimed to replace the global model $\bm{w}_\mathrm{g}$
with a malicious model $\bm{w}_\mathrm{x}$ performing a malicious client selected backdoor task, as in \eqref{equ:model_replacing}.
\begin{align}
	\label{equ:model_replacing}
	\bm{w}_\mathrm{x} = \frac{1}{K} \sum_{k=1}^{K}\bm{w}_k .
\end{align}
As the training process progress, all the clients' model $\bm{w}_k$ converge to $\bm{w}_\mathrm{g}$.
Then, we obtain
\begin{align}
	\bm{w}_\mathrm{x} = \frac{K-1}{K}\bm{w}_\mathrm{g} + \frac{1}{K}\bm{w}_\mathrm{M}, 
\end{align}
where $\bm{w}_\mathrm{M}$ is the model parameter uploaded by the malicious client.
Therefore, the malicious client uploads $\bm{w}_\mathrm{M}$ to replace the global model as follow:
\begin{align}
    \bm{w}_\mathrm{M}= K\bm{w}_\mathrm{x} -(K-1)\bm{w}_\mathrm{g}.
\end{align}
A single shot attack of the malicious client replaces the global model with the malicious model, and the backdoor survives for long rounds without any attacks.
We simply extend this attack to DS-FL; the malicious client sends the logit made by $\bm{w}_\mathrm{x}$, while never updating the model $\bm{w}_\mathrm{x}$.
The model poisoning attack was designed to attack the FL and not the DS-FL.
Thus, this evaluation could not be fair for FL and DS-FL; however, the evaluation reveals the toleration of DS-FL to one of the most powerful attacks designed for FL.

We assumed that the main task was the MNIST task and the backdoor task was the Fashion-MNIST task.
The data distribution over benignant clients was assumed IID.
The malicious client intention was to classify the images of handwritten digits and fashion products of the global model
(e.g., classify images of digit ``0'' and ``T-shirt'' to class ``0'' and images of digit ``3'' and ``Dress'' to class ``3'').
The malicious client included the model trained using the entire MNIST training dataset and Fashion-MNIST training dataset (containing $120{,}000$ images and corresponding labels).
The malicious client performed the model poisoning attack once every five rounds.

\subsection{Results}

\begin{table}[t!]
    \caption{Comparison of communication cost per round in the image classification tasks}
    \centering
    \begin{tabular}{ccc}
        \toprule
        \multirow{2}{*}{Method}&MNIST&Fashion-MNIST\\
        &(smaller model)&(larger model)\\
        \midrule
        Benchmark~1: FL&236.1 MB&1.1 GB\\
        Benchmark~2: FD&40.4 kB&40.4 kB\\
        Proposed: DS-FL&4.0 MB&4.0 MB\\ 
        \bottomrule
    \end{tabular}
    \label{tab:CC/R}
\end{table}
\begin{table}[t]
	\caption{Comparison of communication cost per round for text classification tasks}
	\centering
	\begin{tabular}{ccc}
		\toprule
		\multirow{2}{*}{Method} & IMDb         & Reuters    \\
		                        & (LSTM) & (text-DNN) \\
		\midrule
		Benchmark~1: FL         & 28.6 MB      & 228.8 MB   \\
		Benchmark~2: FD         & 176 B        & 93 kB      \\
		Proposed: DS-FL         & 88 kB        & 2.0 MB     \\
		\bottomrule
	\end{tabular}
	\label{tab:CC/R_text}
\end{table}

\vspace{.3em}\noindent\textbf{Communication cost per round.}\quad
The communication cost per round with FL, FD, and DS-FL are calculated and listed in Tables~\ref{tab:CC/R} and~\ref{tab:CC/R_text} for image classification and text classification tasks, respectively.
As seen in the tables, the communication costs of the proposed DS-FL and the FD benchmark is smaller than that of the FL benchmark and do not depend on the model sizes.
The reason for this result is that the payload size of the logits uploaded in the DS-FL and the FD benchmark is smaller than that in the ML model parameter used in the FL benchmark and does not depend on the number of model parameters.
Furthermore, the communication cost of the FD benchmark is 100 times smaller than that of DS-FL.
The reason is that the number of logits uploaded by a DS-FL client is more than that of the FD.
In the FD benchmark, the clients upload local logits on a per-classy basis, while in the DS-FL, they upload local logits on a per-sample basis in the unlabeled dataset.
However, the proposed DS-FL exhibits a higher classification accuracy than that of the FD benchmark, which is verified in the following results.

\begin{figure}[t!]
    \centering
    \subfloat[MNIST]{
        \includegraphics[width=0.38\textwidth]{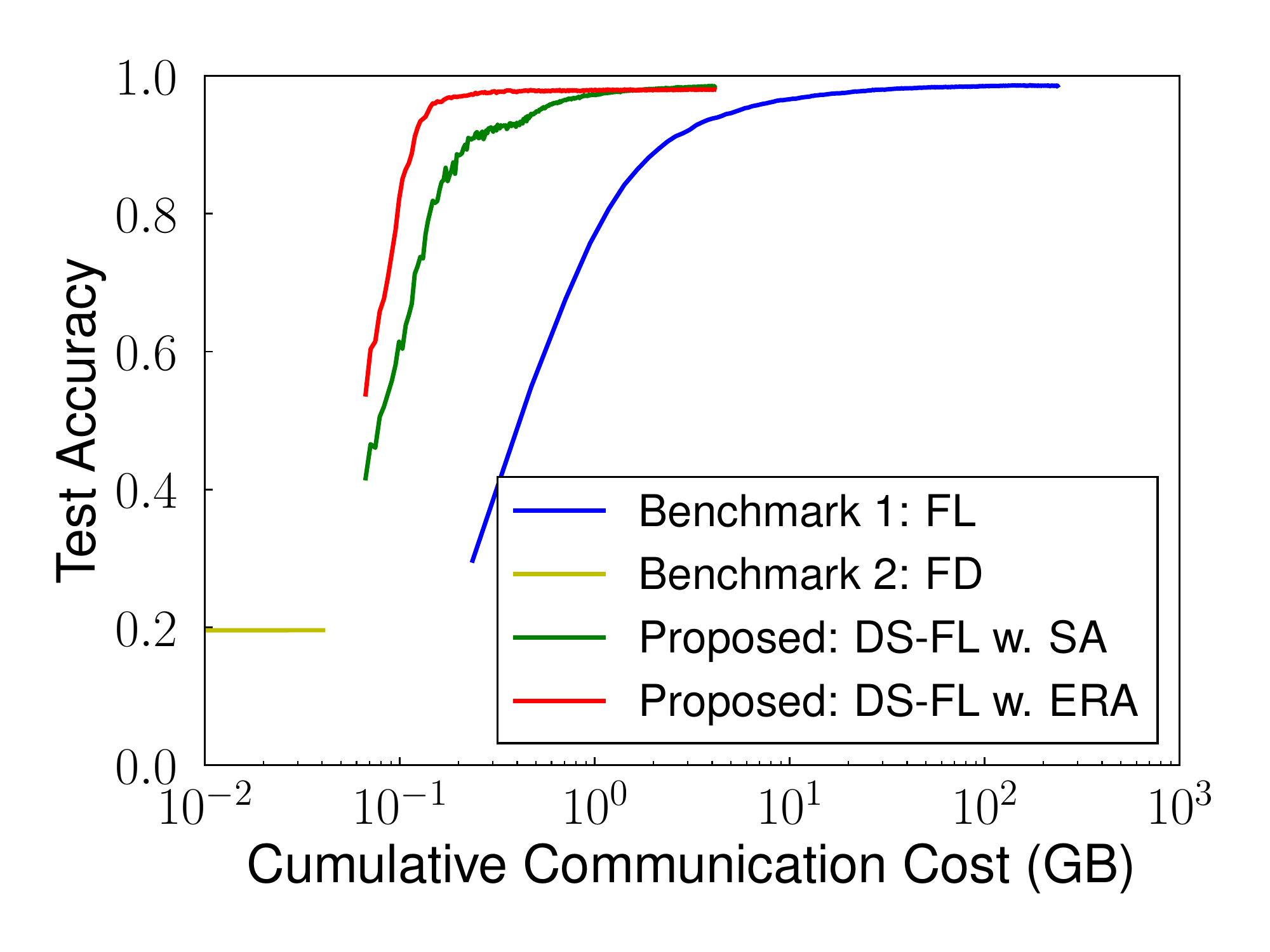}
        \label{fig:MNIST_CCC}
    }\\\vspace{-1em}
    \subfloat[Fashion-MNIST]{
        \includegraphics[width=0.38\textwidth]{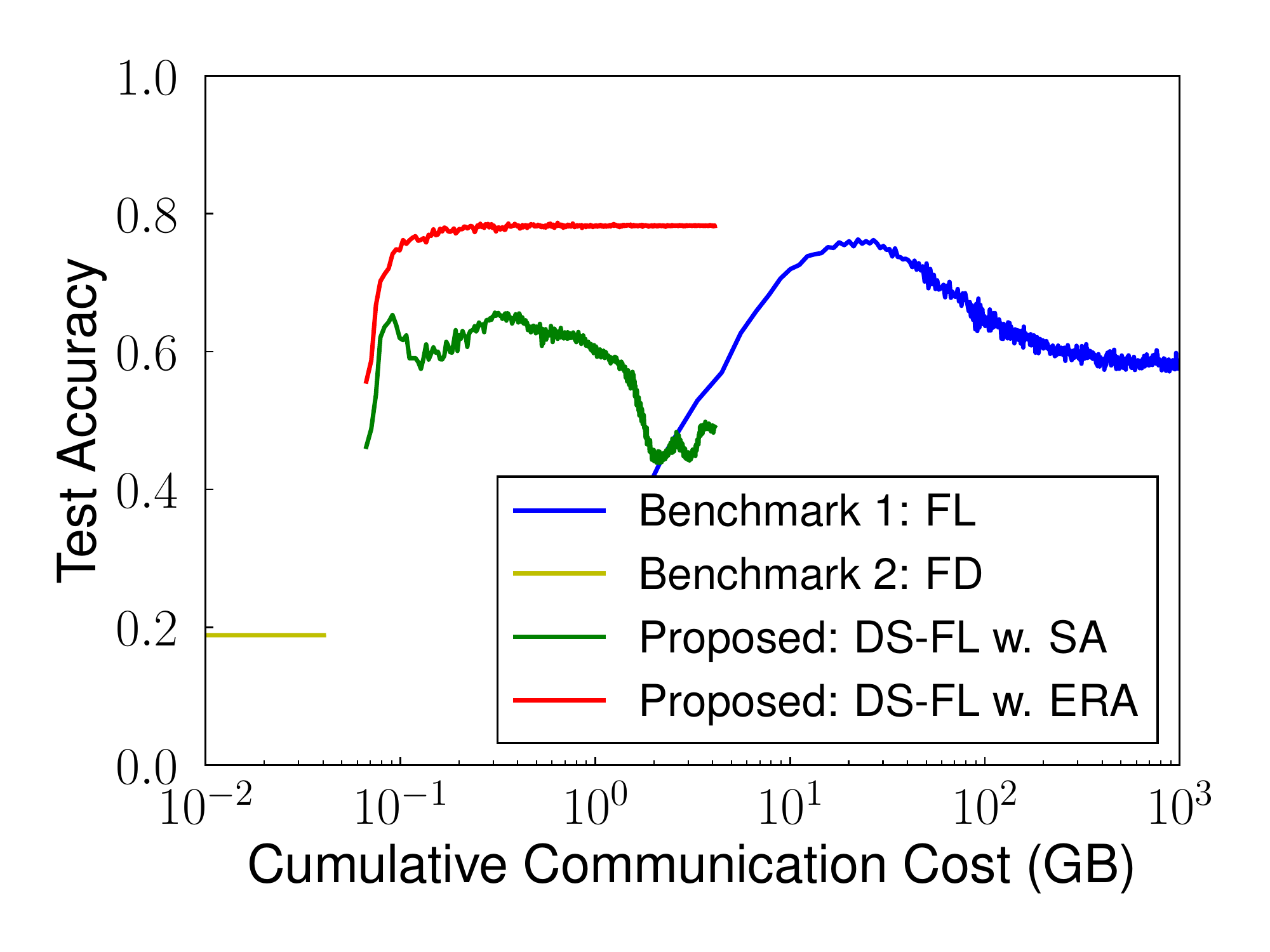}
        \label{fig:F_MNIST_CCC}
    }\\\vspace{-1em}
    \subfloat[IMDb]{
		\includegraphics[width=0.38\textwidth]{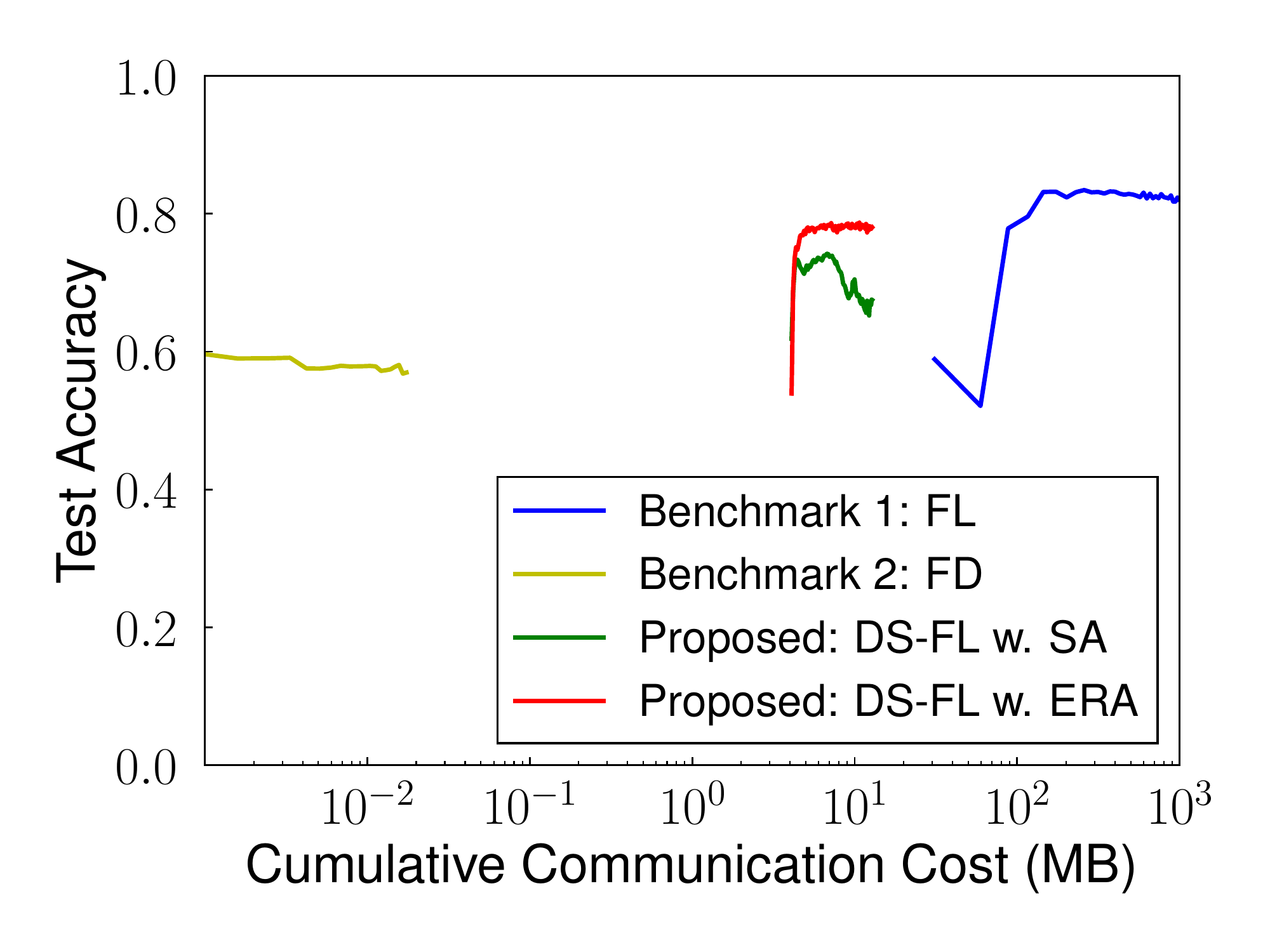}
		\label{fig:IMDb_CCC}
	}\\\vspace{-1em}
	\subfloat[Reuters]{
		\includegraphics[width=0.38\textwidth]{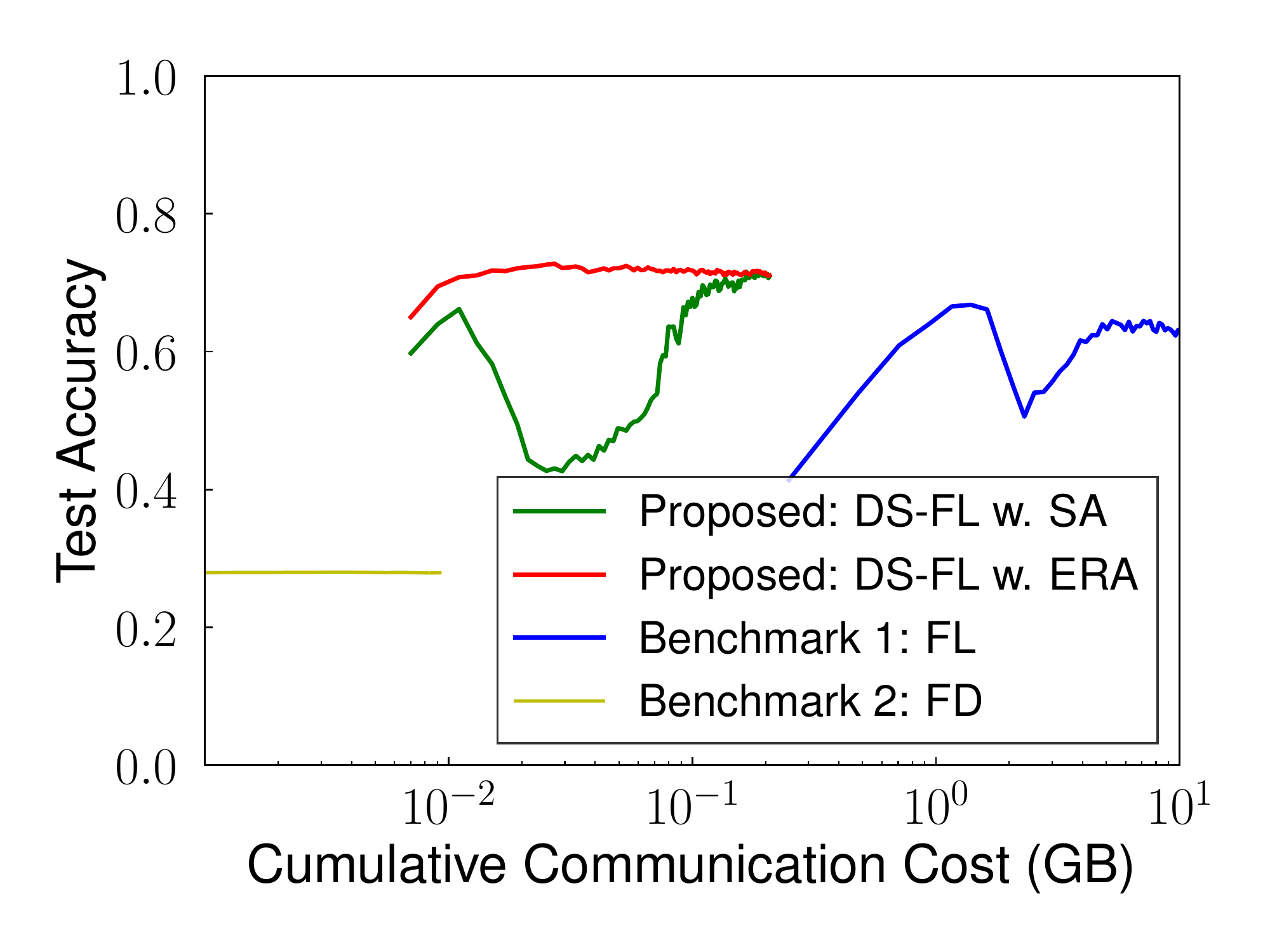}
		\label{fig:Reuters_CCC}
	}
    \caption{Test accuracy vs. cumulative communication costs under non-IID data.
        For MNIST and Fashion-MNIST, the numbers of samples in the local and open unlabeled dataset are $20{,}000$ and $20{,}000$, respectively.}
    \label{fig:CCC}
\end{figure}

\vspace{.3em}\noindent\textbf{Accuracy improvement per communication cost in training.}\quad 
Figs.~\ref{fig:CCC}~(a) and~\ref{fig:CCC}~(b) show the accuracy as a function of the cumulative communication cost 
 for MNIST and Fashion-MNIST, respectively, with $\{I^\mathrm{p},I^\mathrm{o}\} = \{20000 , 20000\}$ and non-IID datasets.
The cumulative communication cost of DS-FL includes an initial cost to distribute the unlabeled data to the clients in addition to the per round cost, while that of other baselines did not include the initial cost.
The initial cost is described as ComU@I in Table~\ref{tab:DDC}. 
In both the FL and DS-FL, as the training processes progress, i.e., the cumulative communication costs to share models or logits increase, and the accuracy improves.
DS-FL performance is evaluated using the global model, which is trained on the server as described in Section 2.3.2.
Meanwhile, the FD accuracy remains approximately $20\%$, which is almost similar to that of a single client.
In both tasks: MNIST and Fashion-MNIST, the proposed DS-FL outperforms the FL (benchmark~1) in terms of cumulative communication cost while achieving comparable accuracy.
As shown in Table~\ref{tab:CC/R}, the results could be due to the communication cost per round of the DS-FL that is lower than that of FL.
If the aggregation methods in DS-FL are compared, the proposed ERA obtains almost the same accuracy as that of the SA baseline, while the cumulative communication cost to the convergence of ERA is smaller than that of SA due to the acceleration effect of ERA.
Based on these results, we can conclude that the DS-FL with the proposed ERA reduces the communication costs substantially while achieving similar performance to the FL benchmark, i.e., ERA accelerates the convergence speed.

Figs.~\ref{fig:CCC}~(c) and~\ref{fig:CCC}~(d) show the accuracy as a function of the cumulative communication cost for IMDb and Reuters.
The results show a similar trend to computer vision tasks, i.e., MNIST and Fashion-MNIST tasks.
In both the FL and DS-FL, as the training processes progress, i.e., the cumulative communication costs to share models or logits increases, and the accuracy improves.
In both tasks, IMDb and Reuters, the proposed DS-FL outperforms the FL (benchmark 1) in terms of cumulative communication cost while achieving comparable accuracy. In contrast, the accuracy of FD is much lower than that of FL, i.e., 23.3\% and 39.0\% lower for IMDb and Reuters, respectively.
If the aggregation methods in DS-FL are compared, the proposed ERA obtains almost the same accuracy as that of the SA baseline, while the cumulative communication cost to the convergence of ERA is smaller than that of SA, due to the acceleration effect of ERA.

Moreover, these results are also verified in Table~\ref{tab:DDC}.
The table lists the cumulative communication costs required to achieve a test classification accuracy of $x\%$, and the highest testing accuracy among the training process, referred to as ComU@$x\%$, and Top-Accuracy, respectively.
In Table~\ref{tab:DDC}, the DS-FL with ERA achieves lower ComU@$x\%$ than that using DS-FL with SA and FL for all cases.
For example, regarding Fashion-MNIST, DS-FL with ERA achieves 99.0\% lower ComU$65\%$ and 99.4\% lower ComU@$75\%$ than FL. Moreover,
regarding Reuters, DS-FL with ERA achieves 99.4\% lower ComU@$65\%$ and higher Top-Accuracy than FL.

\vspace{.3em}\noindent\textbf{Test accuracy comparison.}\quad 
Table~\ref{tab:DDC} lists the highest accuracy for all the training rounds denoted as Top-Accuracy.
The table indicates that the DS-FL with ERA achieves similar or superior Top-Accuracy compared to that of FL and DS-FL with SA.
Comparing DS-FL with ERA to the FL benchmark in the MNIST task, the Top-Accuracy of DS-FL with ERA reaches up to 98.5\% when $I^\mathrm{o}=40{,}000$, which is only 0.2\% lower than that of FL.
In the Fashion-MNIST and Reuters tasks, the Top-Accuracy of the DS-FL with ERA is higher than that of FL for all cases.
In the IMDb task, the Top-Accuracy of the DS-FL with ERA is 4.7\% lower than that of FL, whereas it is 18.6\% higher than that of FD.
Hence, we can again conclude that DS-FL with ERA achieves similar test performance to the FL benchmark while drastically reducing the communication costs.
Comparing the Top-Accuracy of the proposed ERA and SA baseline, SA achieves higher Top-Accuracy relative to ERA for the MNIST task even though the difference becomes smaller as the number of samples in the unlabeled open dataset increases.
Meanwhile, regarding the Fashion-MNIST task, the proposed ERA achieves higher Top-Accuracy than SA, where the difference ranges from 4.0\% to 13.7\%.
Hence, recall that Fashion-MNIST is a more complicated task than MNIST~\cite{F_MNIST}.
In the IMDb and Reuters tasks, the proposed ERA achieves higher accuracy than SA.
These results provide insight into the importance of reducing the entropy global logits, particularly in more complicated tasks, to enhance the DS-FL model performance.

Both FD+FAug~\cite{FD}, which is an advanced method of FD, and DS-FL with ERA deal with non-IID data distribution. Moreover, and FD+FAug outperforms FD.
However, FD+FAug requires clients to upload a part of their labeled data, which does not satisfy the intentions of comparing FL frameworks under the same clients' privacy level.
Thus, comparing DS-FL to other approaches that let clients share their raw data, such as FD+FAug and~\cite{zhao2018federated} is beyond the scope of this study.

\begin{table}[t!]
    \caption{
        Comparison of communication cost and Top-Accuracy with the size of the open dataset $I^\mathrm{o}$.
        ComU@I: the initial cost to distribute open dataset for DS-FL.
        ComU@$x$: the cumulative communicate cost required to achieve a testing classification accuracy of $x$.
        Top-Accuracy: the highest testing accuracy among the training process.
        }
    \centering
    
    \begin{minipage}[tb]{0.49\textwidth}
        \centering
        {(a) \textbf{MNIST}}\\
        {
        \scalebox{0.9}{
        \begin{tabular}{cccccc}
        \\
        \toprule
        \multirow{2}{*}{Method}&\multirow{2}{*}{$I^\mathrm{o}$} & ComU@I& ComU@95\%&ComU@97\% &Top-Acc\\
             && $(\mathrm{GB})$&$(\mathrm{GB})$&$(\mathrm{GB})$&(\%)\\
            \midrule
            Single &\multirow{2}{*}{-} &  \multirow{2}{*}{-}& \multirow{2}{*}{-} & \multirow{2}{*}{-} & \multirow{2}{*}{19.4}\\
            Client\vspace{1pt}\\
            FL&- &  -& 5.67 & 12.5 & 98.7\vspace{1pt} \\
            FD&- &  -& - & - & 19.6 \\
            \midrule
            w.SA&\multirow{2}{*}{$5{,}000$} &  \multirow{2}{*}{0.016}&0.44&0.8&97.6\\
                 w.ERA& &&\textbf{0.09}&-&96.9\\
            \midrule
            w.SA&\multirow{2}{*}{$10{,}000$} & \multirow{2}{*}{0.031}& 0.46 & 0.84 & 98.1 \\
            w.ERA &&&\textbf{0.11}&\textbf{0.18}&97.5 \\
            \midrule
            w.SA&\multirow{2}{*}{$20{,}000$} &\multirow{2}{*}{0.063}& 0.52  & 0.87 & 98.6 \\  
            w.ERA     &&&\textbf{0.14}&\textbf{0.2}&98.1 \\
            \midrule
            w.SA&\multirow{2}{*}{$40{,}000$} &\multirow{2}{*}{0.13}& 0.59& 0.93& 98.7 \\
            w.ERA    & &&\textbf{0.2}&\textbf{0.25}&98.5 \\
        \bottomrule
        \end{tabular}
        }
        }
    \end{minipage}\vspace{1.5em}\\
    \begin{minipage}[tb]{0.49\textwidth}
        \centering
        (b) \textbf{Fashion-MNIST}
        \scalebox{0.9}{
        \begin{tabular}{cccccc}
        \\
        \toprule
        \multirow{2}{*}{Method} &\multirow{2}{*}{$I^\mathrm{o}$}&ComU@I& ComU@65\%&ComU@75\% &Top-Acc\\
            &&$(\mathrm{GB})$&$(\mathrm{GB})$&$(\mathrm{GB})$&(\%)\\
            \midrule
            Single &\multirow{2}{*}{-} &\multirow{2}{*}{-}& \multirow{2}{*}{-} & \multirow{2}{*}{-} &\multirow{2}{*}{18.6} \\
            Client&\vspace{1pt}\\
            FL &-  & -&6.71 & 15.6 & 76.3\vspace{1pt} \\
            FD &- &  -&- & - & 18.9 \\
            \midrule
            w.SA&\multirow{2}{*}{$5{,}000$} & \multirow{2}{*}{0.016}&0.07&-&73.5\\
            w.ERA &&&\textbf{0.03}&\textbf{0.05}&\textbf{77.5}\\
            \midrule
            w.SA&\multirow{2}{*}{$10{,}000$} & \multirow{2}{*}{0.031}&0.22  & - & 68.7\\
            w.ERA &&&\textbf{0.04}&\textbf{0.06}&\textbf{77.1} \\
            \midrule
            w.SA& \multirow{2}{*}{$20{,}000$} &\multirow{2}{*}{0.063}& 0.09 & -& 65.6\\  
            w.ERA &&&\textbf{0.07}&\textbf{0.10}&\textbf{78.7} \\
            \midrule
            w.SA&\multirow{2}{*}{$40{,}000$} & \multirow{2}{*}{0.13}& 0.33& -& 65.3 \\
            w.ERA&&&\textbf{0.14}&\textbf{0.17}&\textbf{79.0} \\
        \bottomrule
        \end{tabular}
        }
    \end{minipage}\vspace{1.5em}\\
    \begin{minipage}[tb]{0.49\textwidth}
		\centering
        {(c) \textbf{IMDb}}\\
        \scalebox{0.9}{
        \centering
		\begin{tabular}{cccccc}
			\\
			\toprule
			\multirow{2}{*}{Method}&$I^\mathrm{o}$&ComU@I& ComU@70\%       & ComU@78\%       & Top-Acc\\
			                        &&$(\mathrm{MB})$& $(\mathrm{MB})$ & $(\mathrm{MB})$ & (\%) \\
			\midrule
            Single           & \multirow{2}{*}{-}& \multirow{2}{*}{-} & \multirow{2}{*}{-}              & \multirow{2}{*}{-}               & \multirow{2}{*}{50.0}  \\
            Client\vspace{1pt}&\\
			FL        &-&- & 87.9            & 116.3           & 83.4\vspace{1pt}  \\
			FD         &-&-& -               & -               & 60.1  \\
			\midrule
			w.SA       &\multirow{2}{*}{$10{,}000$}& \multirow{2}{*}{4.0}            & 4.3             & -               & 74.2  \\
			w.ERA      &&& \textbf{4.2}    & \textbf{5.1}    & 78.7  \\
			\bottomrule
        \end{tabular}
        }
	\end{minipage}\vspace{1.5em}\\
	\begin{minipage}[tb]{0.49\textwidth}
		\centering
        {(d) \textbf{Reuters}}\\
        \scalebox{0.9}{
        \centering
        \begin{tabular}{ccccccc}
            \\
			\toprule
			\multirow{2}{*}{Method}&$I^\mathrm{o}$&ComU@I&  ComU@65\%       & ComU@70\%       & Top-Acc          \\
            && $(\mathrm{MB})$ & $(\mathrm{MB})$ & $(\mathrm{MB})$ &(\%)         \\
			\midrule
            Single            & \multirow{2}{*}{-}& \multirow{2}{*}{-}& \multirow{2}{*}{-}               & \multirow{2}{*}{-}               & \multirow{2}{*}{16.2}          \\
            Client\vspace{1pt}\\
             FL & -& -        & 1163.4          & -               & 66.8\vspace{1pt}        \\
			 FD & -& -        & -               & -               & 28.0          \\
			\midrule
			w.SA & \multirow{2}{*}{$5{,}000$}& \multirow{2}{*}{6.3}                   & 11.0            & 124.4           & 71.7          \\
			w.ERA &&                  & \textbf{7.0}    & \textbf{11.0}   & \textbf{72.8} \\
			\bottomrule
        \end{tabular}
        }
	\end{minipage}
    \label{tab:DDC}
\end{table}

\begin{figure}[t!]
	\centering
	\subfloat[Accuracy]{
		\includegraphics[width=0.44\textwidth]{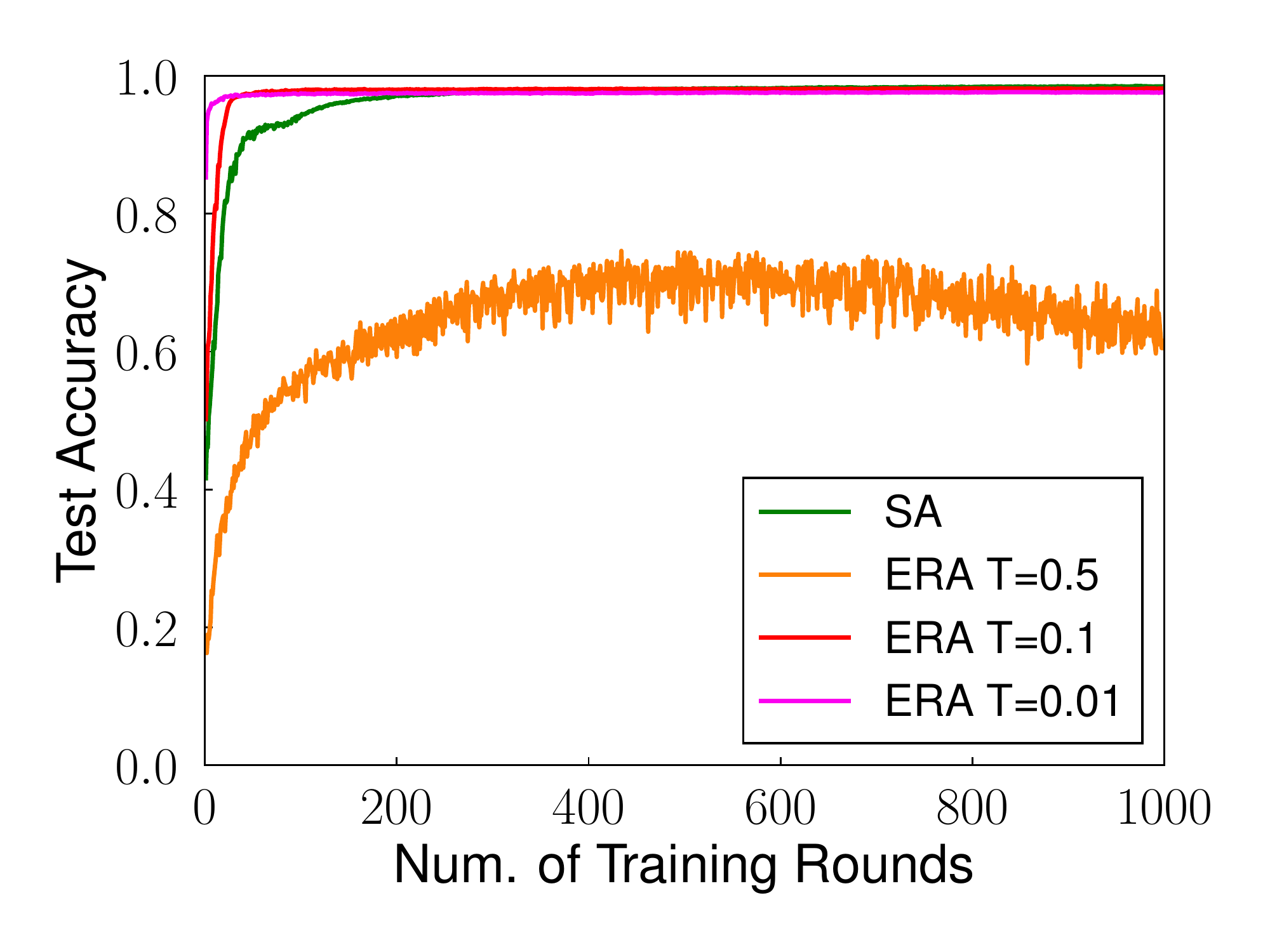}
	}\\\vspace{-1em}
	\subfloat[Entropy of Global logit]{
		\includegraphics[width=0.44\textwidth]{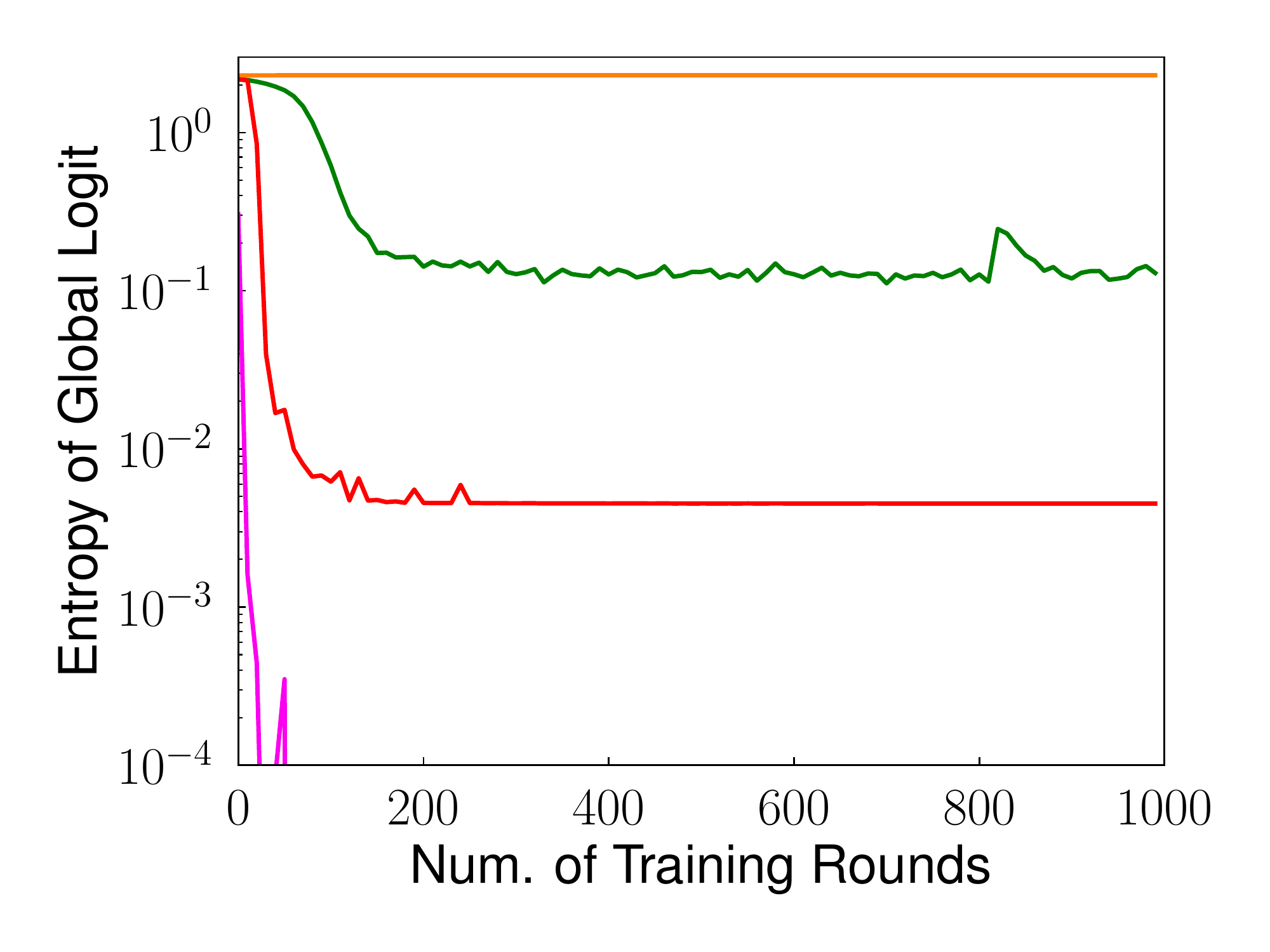}
	}
	\caption{Test accuracy and entropy of global logit vs. training rounds with temperature of ERA, using non-IID and MNIST dataset.}
	\label{fig:ERA_T}
\end{figure}

\vspace{.3em}\noindent\textbf{(Effect of temperture $T$ on entropy reduction aggregation.}\quad 
To evaluate the effect of the temperature on the ERA, we evaluated the performance of ERA according to various $T$ using the MNIST dataset considering non-IID data.
Figs.~\ref{fig:ERA_T}~(a) and~\ref{fig:ERA_T}~(b) show the test accuracy and entropy of global logit as functions of the training round.
When $T=0.5$, the entropy is larger than that of SA, and the training is slower than when considering SA and ERA with smaller $T$.
For $T=0.1$ and $0.01$, the entropy is lower than that of SA, and the training is faster than that when considering SA.
Thus, we note that ERA with a low $T$ accelerates training.

\begin{figure}[t]
    \centering
    \includegraphics[width=0.45\textwidth]{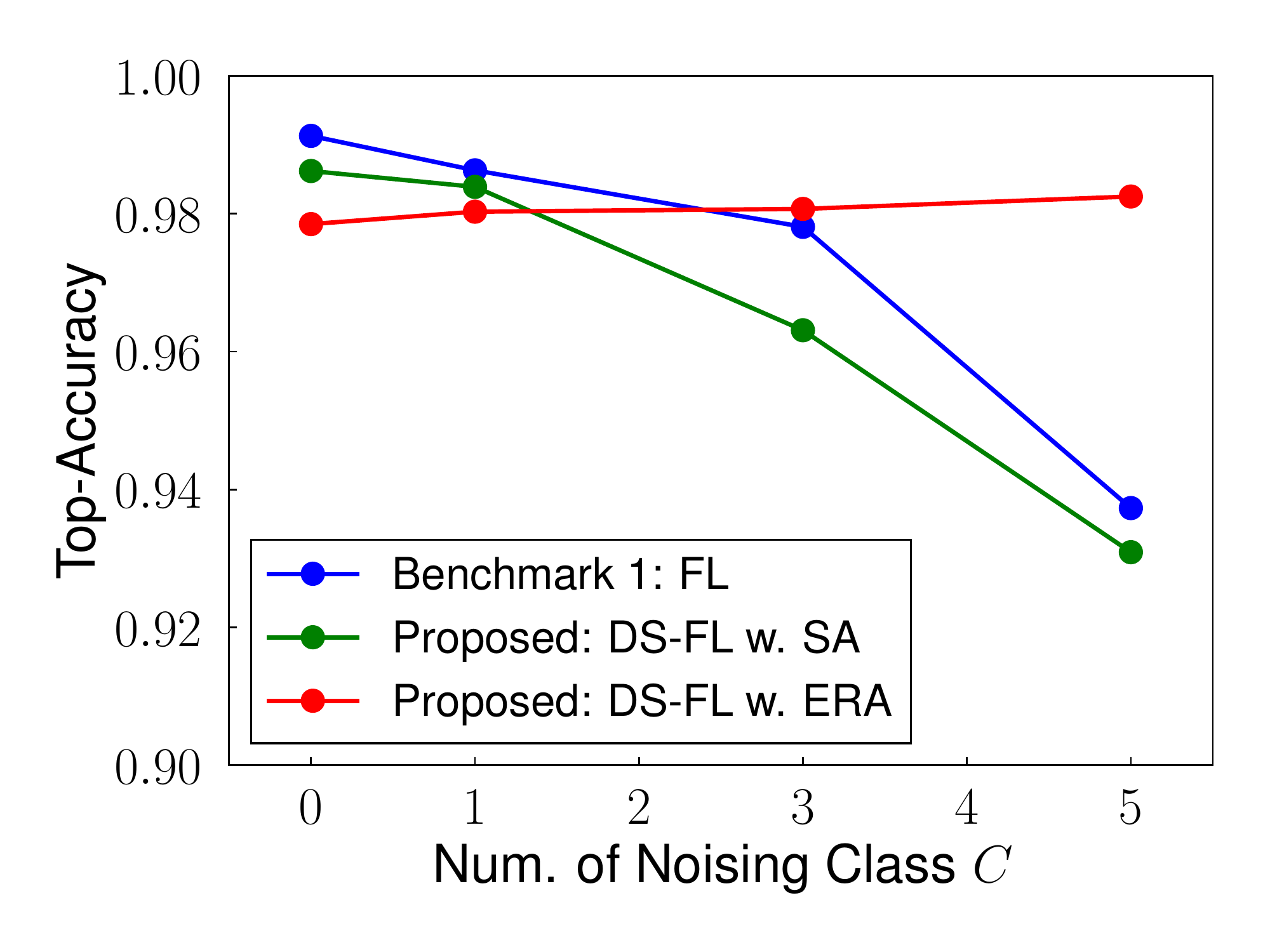}
    \caption{Impact of noisy labels.
    Top-Accuracy as a function of the number of noising classes $C$.
    We use MNIST dataset with $I^\mathrm{p}:I^\mathrm{o} = 40{,}000:20{,}000$.
    The data distribution over client was IID.
    }
    \label{fig:Dist_nonIID}
\end{figure}

\vspace{.3em}\noindent\textbf{Attack robustness of entropy reduction aggregation for noisy labels.}\quad 
Fig.~\ref{fig:Dist_nonIID} shows the Top-Accuracy for MNIST as a function of the number of noising classes $C$.
All the clients hold noisy labeled and properly labeled data samples with a ratio $C : 10-C$.
Regarding the DS-FL with SA and FL, Fig.~\ref{fig:Dist_nonIID} shows that as the noised classes increase, the Top-Accuracy decreases.
However, the DS-FL with ERA maintains the Top-Accuracy, when the noised-classes increase. 
This indicates that the DS-FL with ERA is more robust to IID noising than the FL.
The following section presents an analysis of the global logit entropy to explain the robustness.

\vspace{.3em}\noindent\textbf{Attack robustness of entropy reduction aggregation for noisy open dataset.}\quad 
\begin{figure}[t!]
    \centering
    \includegraphics[width=0.45\textwidth]{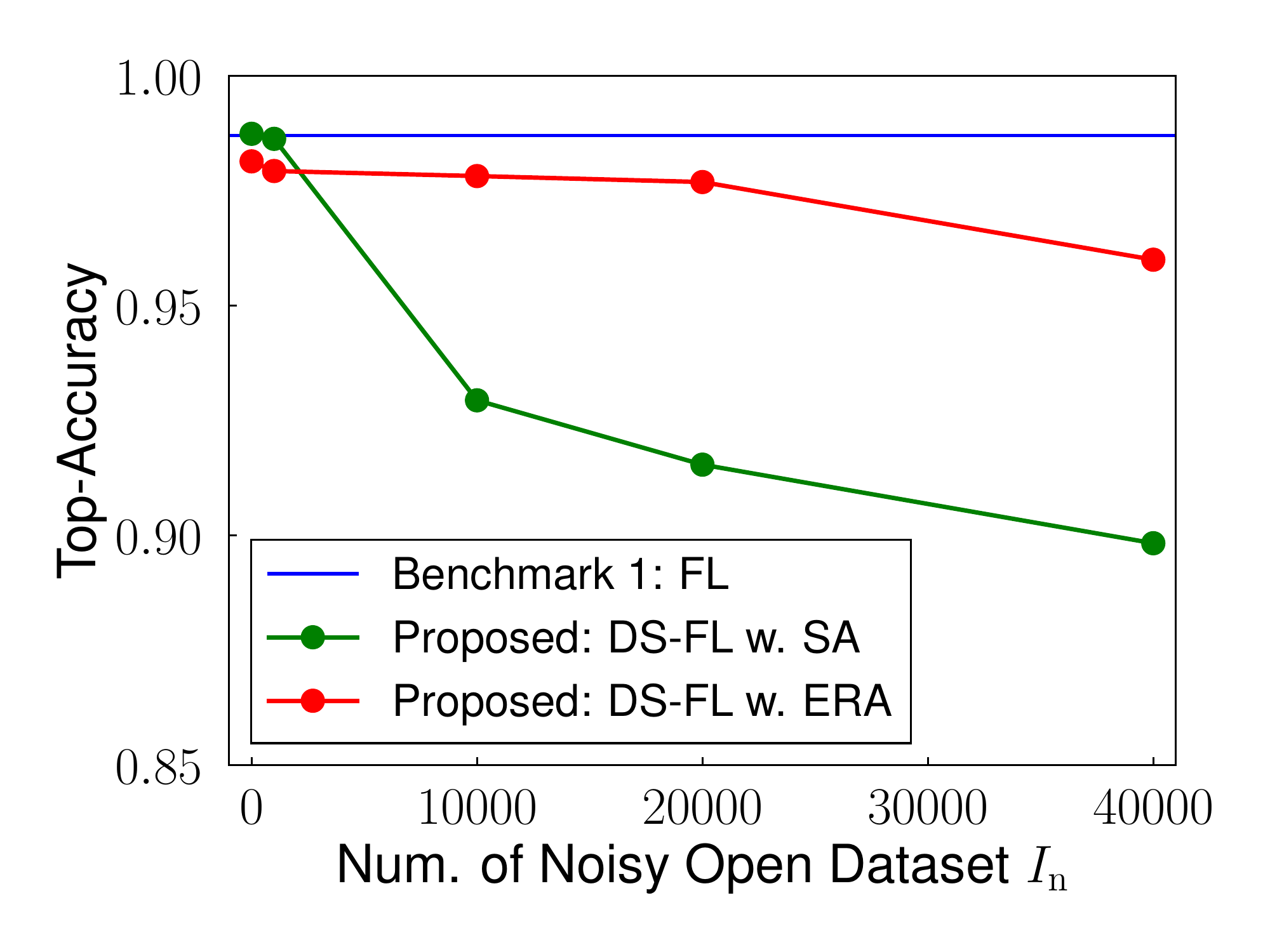}
    \caption{Impact of noisy open dataset.
    Top-Accuracy as a function of the number of noisy open datasets $I_\mathrm{n}$.
    Using MNIST dataset as clean open dataset and Fashion-MNIST dataset as noisy open dataset.
    The size of the clean open dataset was fixed to $20{,}000$.
    The data distribution over client was non-IID.
    }
    \label{fig:Dist_Noisy_Opendata_nonIID}
\end{figure}
Fig.~\ref{fig:Dist_Noisy_Opendata_nonIID} shows the Top-Accuracy for MNIST as a function of the number of noised samples in the open unlabeled dataset.
First, the FL is unaffected by the noisy open data because FL does not use open data.
Overall, as the number of noisy datasets in the open dataset $I^\mathrm{n}$ increases, the Top-Accuracy of DS-FL decreases.
From the perspective of decreasing the Top-Accuracy with a particular number of noisy open datasets from that with $I^\mathrm{n} = 0$ (i.e., the open dataset includes any noisy image) comparing the proposed ERA and the SA baseline, the decrease in Top-Accuracy of ERA is smaller than that of SA.
Hence, we can conclude that the proposed ERA is more robust against a noisy open dataset than the SA baseline.
This result is because the proposed ERA alleviates the increase in the entropy of global logits due to the noisy unlabeled data relative to the SA baseline, as shown in the following section.

\vspace{.3em}\noindent\textbf{Entropy analysis under noisy data attack.}\quad 
\begin{figure}[t!]
    \centering
    \subfloat[
            Entropy of the global logits vs training round with or without noises in the open dataset.
            The open dataset consists of $20{,}000$ benignant images or $20{,}000$ benignant images and $40{,}000$ noisy images.]
        {\includegraphics[width=0.45\textwidth]{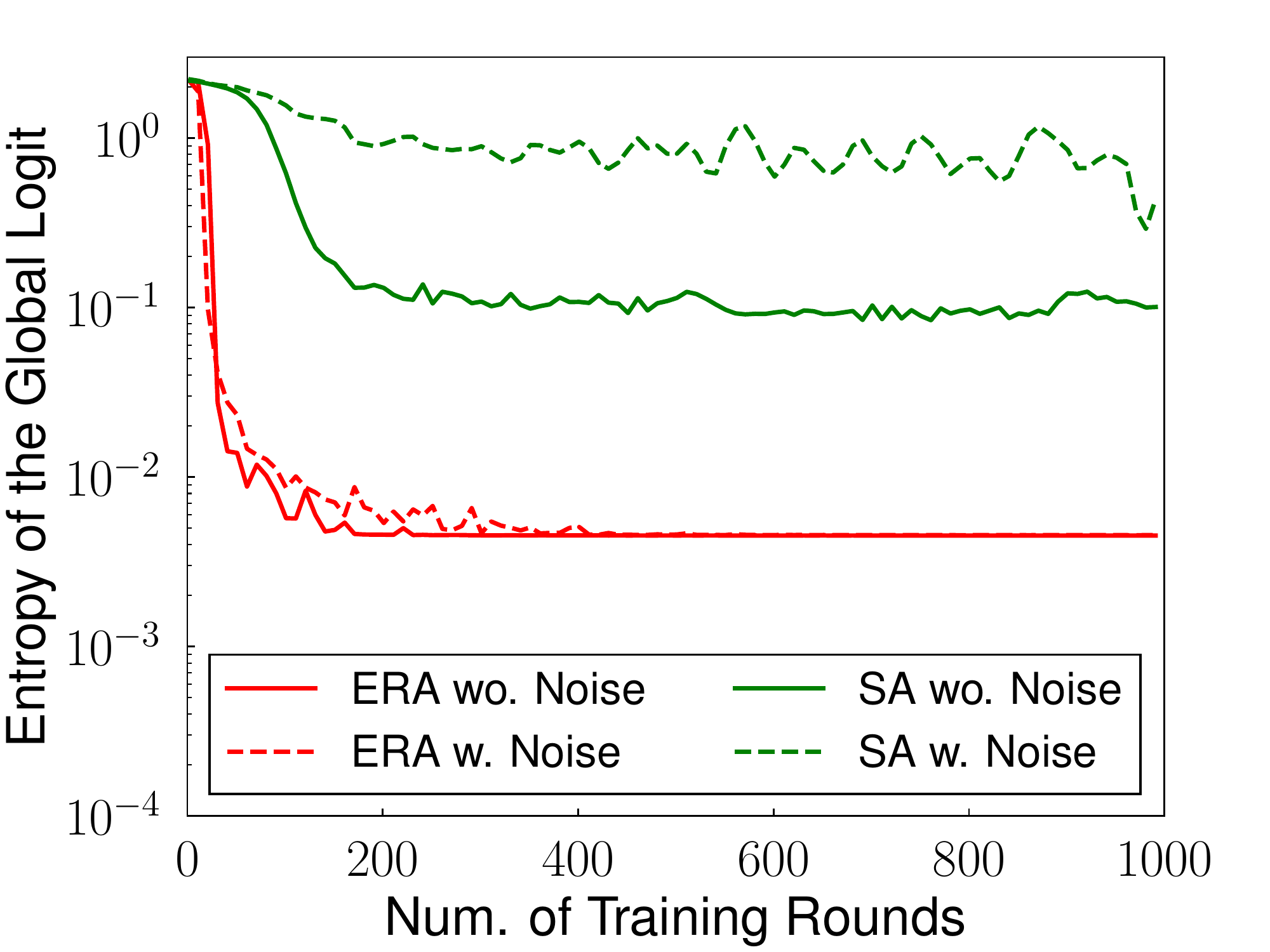}
        \label{fig:EA_noisyopen}
    }
    \\
    \subfloat[
                Entropy of the global logits vs training round with or without noisy label in clients dataset.
                Each client hold 200 noisy labeled dataset and 200 properly labeled dataset.]
            {\includegraphics[width=0.45\textwidth]{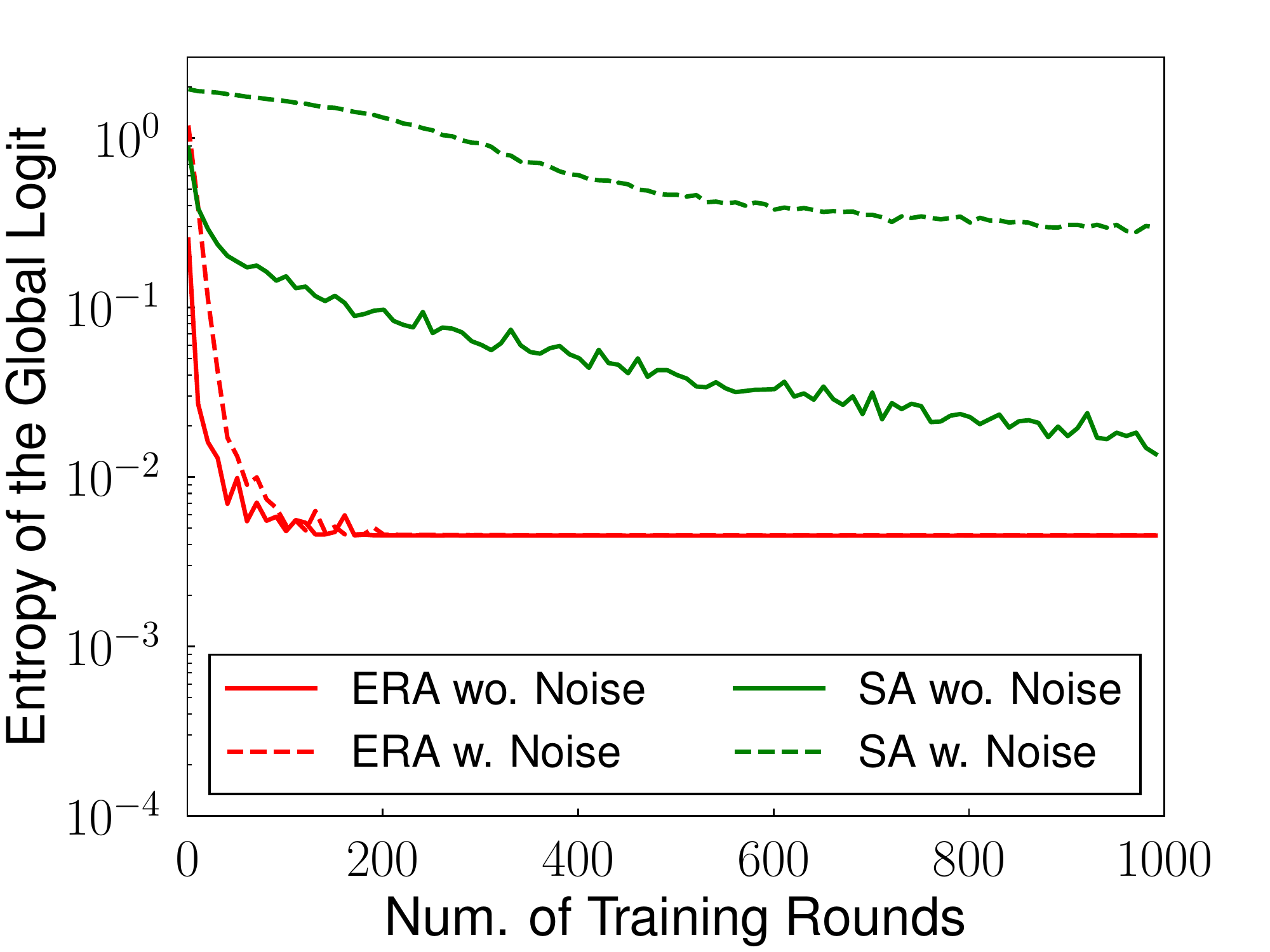}
            \label{fig:EA_noisylabel}
            }
    \caption{
        Entropy of the global logits with or without noises dataset.}
    \label{fig:EA_noise}
\end{figure}
To explain the difference between the proposed ERA and SA baseline, we show in Fig.~\ref{fig:EA_noise} the entropy of the global logits when the datasets include or not noises.
When the dataset includes noises, both SA and ERA entropies become larger relative to that without noises.
Meanwhile, the difference in ERA is smaller than that of SA.
In the SA baseline, the high entropy target vectors are used to train each client's model, making the SA process vulnerable to the noisy open dataset. Simultaneously, the proposed ERA alleviates the increase in the entropy of global logits, leading to high model performance.

\begin{table}[t!]
	\centering
	\caption{Impact of model poisoning attack.
    Test accuracy implies the classification accuracy after 100 rounds.
	}
	\begin{tabular}{cccc}
        \toprule
        \multirow{2}{*}{Method} & Accuracy of & Accuracy of \\
        &main task\%&backdoor task\%\\
		\hline
         FL&98.9&90.4\\
         DS-FL w. SA&97.5&9.6\\
         DS-FL w. ERA&97.9&8.7\\
		\bottomrule
		\\
	\end{tabular}
	\label{tab:malicious}
\end{table}

\vspace{.3em}\noindent\textbf{Attack robustness of Distillaion-Based Semi-Supervised Federated Learning for model poisoning.}\quad
In Table~\ref{tab:malicious}, the attacker achieved its objective in FL, while the attack failed in DS-FL with SA and ERA.
Table~\ref{tab:malicious} shows the test accuracy of the global model after 100 rounds for the main (MNIST) and backdoor tasks (Fashion-MNIST) by the DS-FL and FL.
The objective of the malicious client was to replace the global model with the model achieving high-test accuracy on both the main and the backdoor tasks.
In FL, the global model achieved high-test accuracy on both the main and the backdoor tasks.
This result implies that the objective of the malicious client was achieved.
In contrast, in DS-FL with SA and with ERA, the test accuracy of the global model on the main task was as high as that in FL; however, the accuracy on the backdoor task was much lower than that in FL.
This result implies that the objective of the malicious client was not achieved; thus, the attack failed.
The reason is that DS-FL asks clients to transmit only logit but not ML model parameters,
which prevents the malicious client from the attack that needs altering the uploading model parameters.
 
\section{Conclusion}
\label{sec:conclusion}
We proposed a cooperative learning method, named DS-FL, designed to be scalable according to model sizes in terms of communication efficiency while achieving similar accuracy to benchmarks FL algorithms.
The fundamental idea of the proposed DS-FL was the model output exchange for an unlabeled open dataset.
Additionally, we proposed a logit aggregation method for the DS-FL, which aimed to accelerate the training process and enhance robustness under the non-IID data.
The simulations showed that the proposed DS-FL method outperformed the benchmark method FL in terms of communication cost and robustness while achieving similar or superior accuracy to that of the FL.
Moreover, the experimental results showed that the DS-FL with ER was more communication efficient and robust than the DS-FL with SA.
To explain the performance of the proposed methods, we analyzed the experimental results from the perspective of entropy.
Additionally, the impact of the open dataset volume was evaluated.

The future works will include developing the logit aggregation method, considering the individual device characteristics.
For example, enhancing the impact of the logits uploaded by the reliable or high-performance client with respect to the global logit.
However, how to evaluate the reliability of the clients, and how to control the impact of the uploaded logit are unknown.
Another interesting direction is leveraging the logits had uploaded in the past round.
In this work, the server and the clients used the logits uploaded at the current round.
However, it might be useful to note that the logits had been uploaded in the past rounds.
Moreover, another direction of future work is to design an FL framework performing under non-IID data distributions and unbalanced and massively distributed data while achieving communication costs scalability.
To evaluate the FL framework under unbalanced and massively distributed data, the benchmarking framework for FL, LEAF~\cite{caldas2018leaf}, will be helpful.

\section*{Acknowledgments}
This work was supported in part by JSPS KAKENHI Grant Numbers JP17H03266, JP18K13757, JST PRESTO Grant Number JPMJPR2035, and the KDDI Foundation.

\bibliographystyle{IEEEtran}
\bibliography{IEEEabrv,main}

\begin{IEEEbiography}
[{\includegraphics[width=1in, height=1.25in, clip, keepaspectratio]{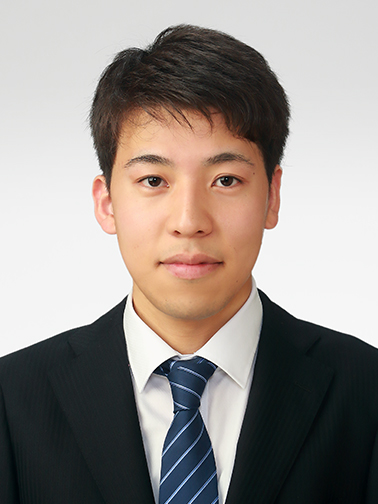}}]
{Sohei~Itahara}
received the B.E. degree in electrical and electronic engineering from Kyoto University in 2020.
He is currently studying toward the M.I. degree at the Graduate School of Informatics, Kyoto University.
He is a student member of the IEEE.
\end{IEEEbiography}
    
\begin{IEEEbiography}
    [{\includegraphics[width=1in, height=1.25in, clip, keepaspectratio]{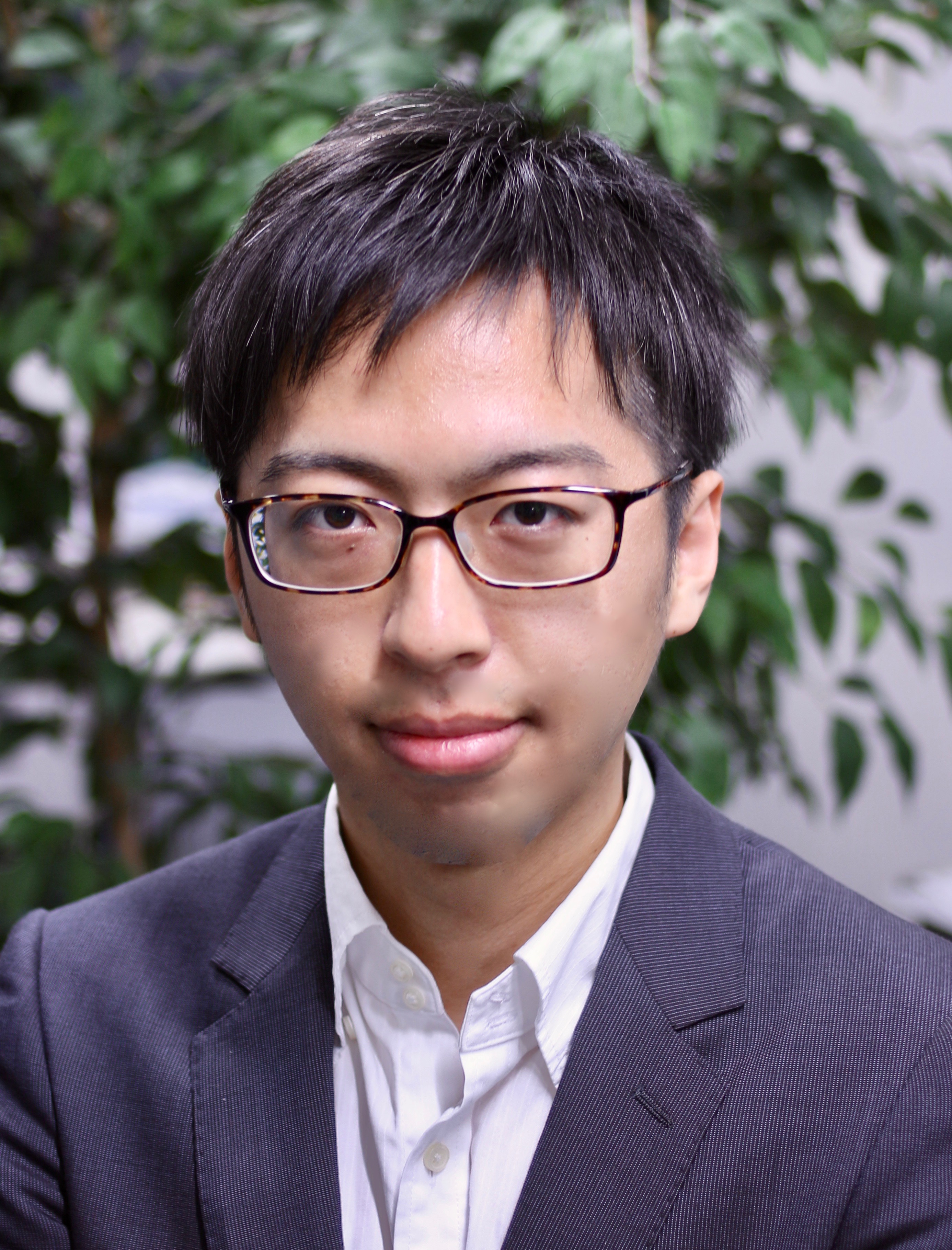}}]
    {Takayuki~Nishio}
    has been an associate professor in the School of Engineering, Tokyo Institute of Technology, Japan, since 2020. He received the B.E.\ degree in electrical and electronic engineering and the master's and Ph.D.\ degrees in informatics from Kyoto University in 2010, 2012, and 2013, respectively. He had been an assistant professor in the Graduate School of Informatics, Kyoto University from 2013 to 2020. From 2016 to 2017, he was a visiting researcher in Wireless Information Network Laboratory (WINLAB), Rutgers University, United States. His current research interests include machine learning-based network control, machine learning in wireless networks, and heterogeneous resource management.
\end{IEEEbiography}
        
\begin{IEEEbiography}[{\includegraphics[width=1in, height=1.25in, clip, keepaspectratio]{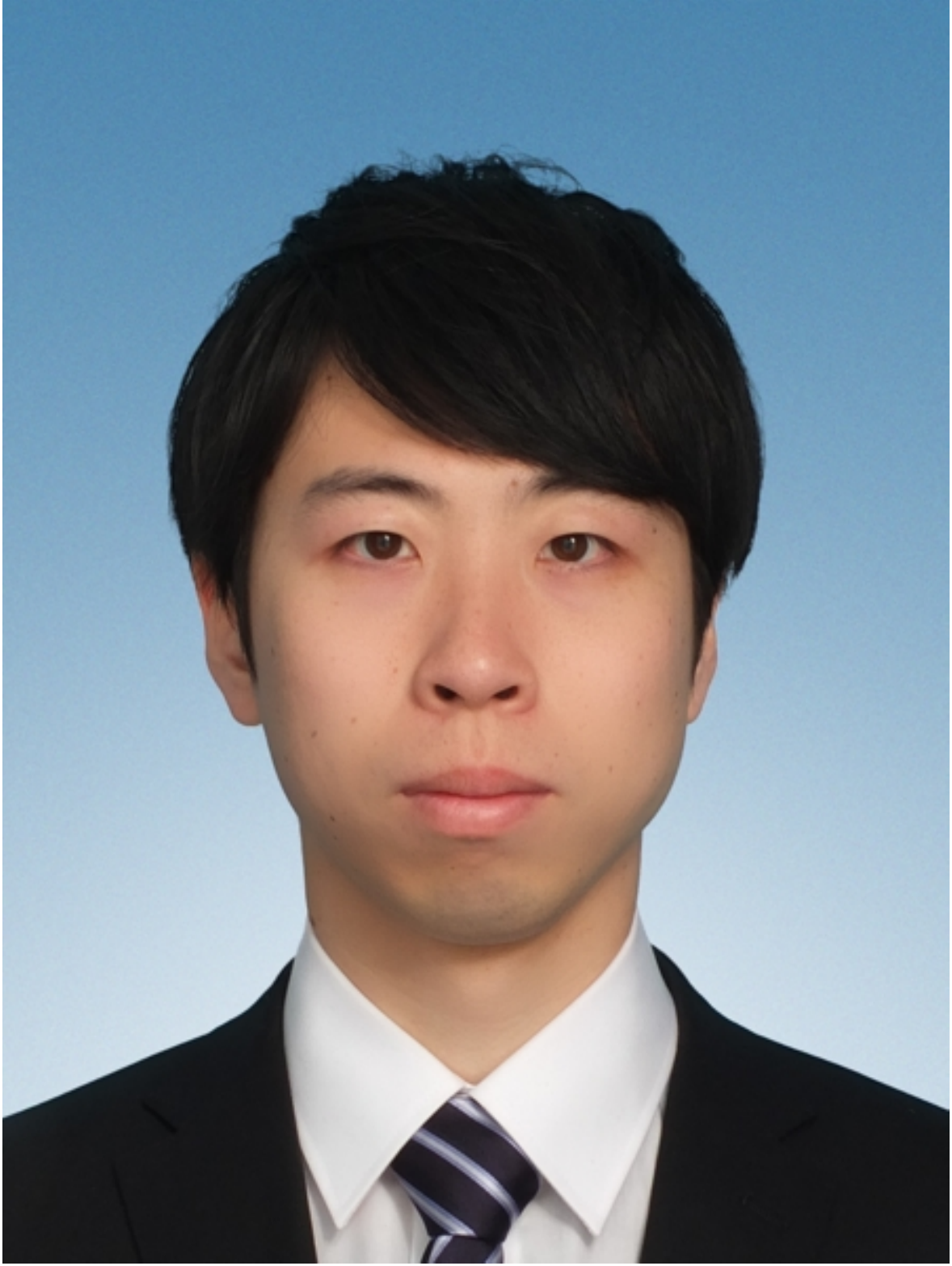}}]{Yusuke~Koda}
    received the B.E. degree in electrical and electronic engineering from Kyoto University in 2016 and the M.E. degree at the Graduate School of Informatics from Kyoto University in 2018.
    In 2019, he visited Centre for Wireless Communications, University of Oulu, Finland to conduct collaborative research.
    He is currently studying toward the Ph.D. degree at the Graduate School of Informatics from Kyoto University.
    He was a Recipient of the Nokia Foundation Centennial Scholarship in 2019.
    He received the VTS Japan Young Researcher's Encouragement Award in 2017.
    He is a member of the IEICE and a member of the IEEE.
\end{IEEEbiography}

\begin{IEEEbiography}
    [{\includegraphics[width=1in, height=1.25in, clip, keepaspectratio]{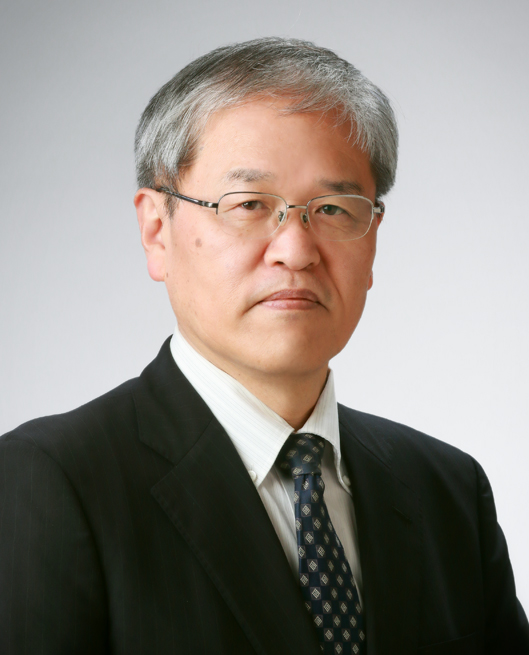}}]
    {Masahiro~Morikura}
    received B.E., M.E. and Ph.D. degree in electronic engineering from Kyoto University, Kyoto, Japan in 1979, 1981 and 1991, respectively. He joined NTT in 1981, where he was engaged in the research and development of TDMA equipment for satellite communications.  From 1988 to 1989, he was with the communications Research Centre, Canada as a guest scientist. From 1997 to 2002, he was active in standardization of the IEEE802.11a based wireless LAN. He received Paper Award and Achievement Award from the IEICE in 2000 and 2006, respectively. He also received Education, Culture, Sports, Science and Technology Minister Award in 2007 and the Medal of Honor with Purple Ribbon from Japan's Cabinet Office in 2015. Dr. Morikura is now a professor of the Graduate School of Informatics, Kyoto University. He is a Fellow of the IEICE. 
\end{IEEEbiography}

\begin{IEEEbiography}
    [{\includegraphics[width=1in, height=1.25in, clip, keepaspectratio]{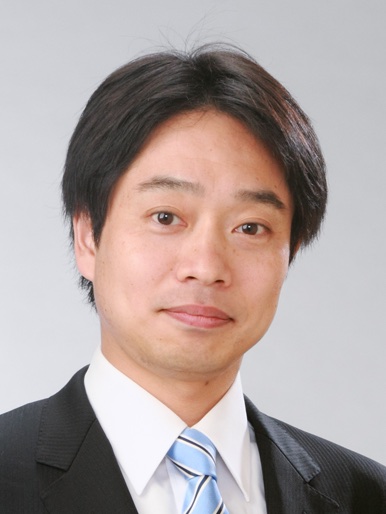}}]
    {Koji~Yamamoto}
    received the B.E. degree in electrical and electronic engineering from Kyoto University in 2002, and the M.E. and Ph.D. degrees in Informatics from Kyoto University in 2004 and 2005, respectively.
    From 2004 to 2005, he was a research fellow of the Japan Society for the Promotion of Science (JSPS).
    Since 2005, he has been with the Graduate School of Informatics, Kyoto University, where he is currently an associate professor.
    From 2008 to 2009, he was a visiting researcher at Wireless@KTH, Royal Institute of Technology (KTH) in Sweden.
    He serves as an editor of IEEE Wireless Communications Letters from 2017 and the Track Co-Chairs of APCC 2017 and CCNC 2018.
    His research interests include radio resource management and applications of game theory.
    He received the PIMRC 2004 Best Student Paper Award in 2004, the Ericsson Young Scientist Award in 2006.
    He also received the Young Researcher's Award, the Paper Award, SUEMATSU-Yasuharu Award from the IEICE of Japan in 2008, 2011, and 2016, respectively, and IEEE Kansai Section GOLD Award in 2012.
\end{IEEEbiography}
\end{document}